\documentclass[a4paper,twocolumn,pra,superscriptaddress,amsmath,amssymb,mathrsfs,showpacs,nofootinbib]{revtex4-1}
\usepackage{graphicx}
\usepackage{epstopdf}
\usepackage[table]{xcolor}     
\usepackage{braket}
\usepackage[colorlinks]{hyperref} 

\usepackage{bbold}

\hypersetup{%
linkcolor=blue,
citecolor=blue,
filecolor=black,
menucolor=black,
urlcolor=black
}
\usepackage[customcolors,norndcorners]{hf-tikz}
\hfsetfillcolor{gray!10}
\hfsetbordercolor{gray!50}

\newcommand\norm[1]{\lVert#1\rVert}
\newcommand\tr[1]{\text{tr}\left[#1\right]}
\newtheorem{theorem}{Theorem}

\begin{document}
\title{Optimized Sampling of Mixed-State Observables}
\affiliation{Theoretische Physik, Universit\"at Kassel,\ Heinrich Plett-Stra\ss e 40,\ 34132 
Kassel, Germany}  
\author{Marec W. Heger} 
\author{Christiane P. Koch}
\author{Daniel M. Reich}
\email{daniel.reich@uni-kassel.de}

\begin{abstract}
Quantum dynamical simulations of statistical ensembles pose a significant computational challenge due to the fact that mixed states need to be represented. 
If the underlying dynamics is fully unitary, for example in ultrafast coherent control  at finite temperatures, one approach to approximate time-dependent observables 
is to sample the density operator 
by solving the Schr\"{o}dinger equation for a set of wave functions with randomized phases. We show that, on average, random-phase wave functions perform well for ensembles with high mixedness, whereas at higher purities a deterministic sampling of the energetically lowest-lying eigenstates becomes superior. We prove that minimization of the worst-case error for computing arbitrary observables is uniquely attained by eigenstate-based sampling.
We show that this error can be used to form a qualitative estimate of the set of ensemble purities for which the sampling performance of the eigenstate-based approach is superior to random-phase wave functions. 
Furthermore we present refinements to both schemes which remove redundant information from the sampling procedure to accelerate their convergence. Finally, we point out how the structure of low-rank observables can be exploited to further improve eigenstate-based sampling schemes.
\end{abstract}

\maketitle

\section{Introduction}
\label{sec:introduction}

In many molecular and condensed matter systems the efficient time propagation of statistical ensembles is crucial to properly model quantum dynamics under realistic conditions. Such incoherent states are most commonly described by density matrices which incorporate the concept of classical mixtures on top of quantum coherences~\cite{Breuer2002}. However, the numerical treatment of density matrices proves to be challenging due to the requirement to store $\mathcal{O}(N^2)$ complex numbers for a general representation, where $N$ is the underlying Hilbert space dimension. Conversely, pure states in Hilbert space only require the storage of at most $\mathcal{O}(N)$ complex entries. This issue is further exacerbated by the fact that, on top of the increased memory requirements, the computational effort is substantially larger, too. In the most general case, time propagation via solving the Schr\"odinger equation for pure states scales as $\mathcal{O}(N^2)$ and grows to $\mathcal{O}(N^4)$ when solving, e.g., the Liouville equation for density matrices.

A promising approach to reduce computational complexity is to find an effective description in Hilbert space. Methods to treat dissipative systems on the level of coherent states include, for example, Monte Carlo wave function techniques~\cite{Dalibard1992,Gardiner1992}, Keldysh contour methods~\cite{Sieberer2016}, and the so-called Surrogate Hamiltonian~\cite{Koch2002,Koch2003}. An interesting subclass of problems is given by mixed states which undergo coherent evolution. If no dissipative processes are present on the timescale of the dynamics, solving the time-dependent Schr\"{o}dinger equation for a complete set of Hilbert space states yields an exact representation of the system's time evolution, reducing the computational complexity from $\mathcal{O}(N^4)$ to $\mathcal{O}(N^3)$~\cite{Koch2006}. This raises the question whether a further reduction can be achieved by preselecting an incomplete set of Hilbert space states.

The search for approaches in which one only considers a small subset of the total Hilbert space is motivated by the so-called eigenstate thermalization hypothesis~\cite{Deutsch1991,Srednicki1994,Rigol2008}. It states that, under certain conditions, the behavior of a complete statistical ensemble can be reproduced by individual energy eigenstates. As such one can expect that ensemble observables can already be approximated by considering only a small set of initial wave functions. A popular idea in this direction is the use of so-called random-phase thermal wave functions~\cite{GelmanCPL2003}. This method rests on the observation that while individual Hilbert space states can properly represent populations of an incoherent ensemble, they usually contain excess coherences. However, by randomizing the phases of these coherences through using a set of random-phase thermal wave functions the superfluous contributions can be averaged out, thus restoring a proper description of the initial state.

Such a stochastic approach has already been proven to yield well-converged results with a comparatively small number of random-phase realizations in applications such as photoassociation of Mg$_2$ dimers~\cite{RybakPRL2011,AmaranJCP2013,LevinPRL2015} and laser-induced rotation of SO$_2$ molecules~\cite{Kallush2015}. However, its asymptotic behavior is rather poor, showing the well-known statistical $\sim \frac{1}{\sqrt{K}}$ tail where $K$ is the number of realizations employed. Moreover, in contrast to using an orthonormal basis of Hilbert space, the initial ensemble is not fully reobtained even when $K$ is equal to the Hilbert space dimension. This poses the question whether an approach using an orthogonal set of wave functions is superior to random-phase thermal wave functions in computing time-dependent expectation values.

Here we provide a theorem which uniquely identifies the best set of Hilbert space states to limit the worst-case approximation error of arbitrary time-dependent observables in mixed ensembles. In addition to the worst-case estimate, we
analyze the average behavior of this 
state sample for a set
of physical observables and compare it to the random-phase approach. 
Finally, we show that prior information about an observable, if it is available, may be used to design even faster converging state sets to approximate the mixed-state expectation value of this observable. 

The paper is organized as follows. Section~\ref{sec:random_phase} introduces the concept of wave-function sampling to simulate time-dependent ensemble expectation values. Section~\ref{sec:opt_sampling_arb_observable} introduces an error measure for the performance of a given sampling scheme and presents our central theorem on how to minimize the worst-case estimation error for arbitrary observables. Section~\ref{sec:sampling_enhancements} proposes two modifications to the previously introduced sampling methods that lead to faster convergence of the average error. 
In Section~\ref{sec:app_to_random_observables} we apply the two sampling protocols to a one-dimensional spin chain and discuss how they behave on average, with a particular emphasis on the dependence on ensemble purity. Section~\ref{sec:opt_sampling_low_rank_observable} explores in how far prior information can be used to further optimize the sampling procedure. Finally, Section~\ref{sec:conclusions} concludes.

\section{Random-phase thermal wave functions}
\label{sec:random_phase}

\subsection{Eigenstate-based Sampling}
\label{subsec:eigenstate_sampling}

We start by considering an initial state in thermal equilibrium, i.e., a canonical ensemble and generalize to arbitrary mixed states in the following. The corresponding density operator is given by
\begin{equation}
\hat{\rho}_\beta=\frac{e^{-\beta\hat{H}_0}}{Z}\,,
\end{equation} 
where $Z=\tr{e^{-\beta\hat{H}_0}}$ is the partition function, $\hat{H}_0\equiv\hat{H}(t=0)$ the system Hamiltonian at initial time, and $\beta=1/(k_BT)$ the inverse temperature. We denote the eigenbasis of $\hat{H}_0$ as $\{\ket{E_n}\}_{n=1,\dots,N}$ with eigenenergies $\{E_n\}_{n=1,\dots,N}$. This diagonalizes the density matrix,
\begin{equation}
\label{Eq:rho_true_in_energy_eigenbasis}
\hat{\rho}_\beta=\frac{1}{Z}\sum_{n=1}^{N}e^{-\beta E_n}\ket{E_{n}}\bra{E_{n}}\,,
\end{equation} 
with $N$ the Hilbert space dimension. Assuming coherent system dynamics, time propagation is mediated by the unitary time evolution operator, $\hat{U}(t)$, generated by the time-dependent Hamiltonian $\hat{H(t)}$. As a consequence, expectation values of observables at a certain time, $\braket{\hat{A}}_{\beta}$, can be obtained via solving the time-dependent Schr{\"o}dinger equation (instead of the time-dependent Liouville equation for the full density matrix). Specifically, one can write 
\begin{eqnarray}
\braket{\hat{A}}_\beta&=&\text{tr}[\hat{A}\hat{U}(t)\hat{\rho}_\beta\hat{U}^\dagger(t)]=\text{tr}[\hat{U}^\dagger(t)\hat{A}\hat{U}(t)\hat{\rho}_\beta]\nonumber\\
&=&\frac{1}{Z}\sum_{n=1}^{N}\bra{E_{n}}\hat{U}^\dagger(t)\hat{A}\hat{U}(t)e^{-\beta E_n}\ket{E_{n}}\nonumber\\
&=&\sum_{n=1}^{N}p_n\bra{E_{n}}\hat{U}^\dagger(t)\hat{A}\hat{U}(t)\ket{E_{n}},\label{Eq:expectation_value_in_energy_basis}
\end{eqnarray}
with $p_n\equiv\frac{1}{Z}e^{-\beta E_n}$. As a result, the expectation value of an arbitrary observable can be computed exactly by Hilbert space propagation of a complete orthonormal basis. However, for systems with large Hilbert space dimension, this quickly becomes infeasible. This phenomenon is called the curse of dimensionality~\cite{Bellman1957} leading to exponential scaling for each additional degree of freedom. One approach to reduce the numerical effort is truncation of Eq.~\eqref{Eq:expectation_value_in_energy_basis} once the weights $p_n$ become sufficiently small~\cite{Koch2006}, i.e. 
\begin{equation}
\braket{\hat{A}}_\beta\approx\sum_{n=1}^{K}p_n\bra{E_{n}}\hat{U}^\dagger(t)\hat{A}\hat{U}(t)\ket{E_{n}}\label{Eq:eigenbasis_sampling}
\end{equation}
This approach is sensible for very cold ensembles, where the weights quickly drop off for excited states due to the Boltzmann factor $e^{-\beta E_n}$. For moderately warm ensembles the distribution broadens and, to our knowledge, no feasible way to predict the incurred error in the observable for a given cutoff point has been derived.

\subsection{Thermal wave function sampling}
\label{subsec:thermal_sampling}
Instead of using a truncated orthonormal eigenbasis, Gelman and Kosloff proposed random-phase thermal wave functions to approximate $\hat{\rho}_\beta$ and thereby compute time-dependent observables $\braket{\hat{A}}_{\beta}$~\cite{GelmanCPL2003}. In a first step, a set of $K$ non-orthogonal states, $\{\ket{\Phi_k}\}_{k=1,\dots,K}$, is constructed from some orthonormal basis $\{\ket{\phi_j}\}_{j=1,\dots,N}$,
\begin{equation}
\ket{\Phi_k(\vec{\theta}^k)}=\frac{1}{\sqrt{N}}\sum_{j=1}^Ne^{i\theta_j^k}\ket{\phi_j}\,.
\label{Eq:thermal_linear_combination}
\end{equation}
Each $\ket{\Phi_k}$ uses randomly chosen phases $\theta_j^k\in[0,2\pi[$. The dyadic product $\ket{\Phi_k}\bra{\Phi_k}$ is called realization,
\begin{equation}
\ket{\Phi_k}\bra{\Phi_k}=\frac{1}{N}\sum_{j,j'=1}^{N}e^{i(\theta_j^k-\theta_{j'}^k)}\ket{\phi_j}\bra{\phi_{j'}}\,.
\end{equation}
By averaging an infinite number of realizations a resolution of the identity, $\hat{\mathbb{1}}$, is obtained,
\begin{equation}
\hat{\mathbb{1}}=\lim\limits_{K\rightarrow\infty}\left(\frac{N}{K}\sum_{k=1}^{K}\ket{\Phi_k}\bra{\Phi_k}\right)\,.\label{Eq:identity}
\end{equation}
Using Eq.~\eqref{Eq:identity}, $\hat{\rho}_\beta$ can be expressed in random-phase thermal wave functions,
\begin{eqnarray}
\hat{\rho}_\beta&=&\frac{1}{Z}e^{-(\beta/2)\hat{H}_0}\hat{\mathbb{1}}e^{-(\beta/2)\hat{H}_0}\nonumber\\
&=&\lim\limits_{K\rightarrow\infty}\frac{N}{ZK}\sum_{k=1}^{K}e^{-(\beta/2)\hat{H}_0}\ket{\Phi_k}\bra{\Phi_k}e^{-(\beta/2)\hat{H}_0}\nonumber\\
&=&\lim\limits_{K\rightarrow\infty}\frac{N}{K}\sum_{k=1}^{K}\Ket{\Phi_k\left(\frac{\beta}{2},\vec{\theta}^k\right)}\Bra{\Phi_k\left(\frac{\beta}{2},\vec{\theta}^k\right)}\,,\label{Eq:rho_thermal}
\end{eqnarray}  
with unnormalized $\Ket{\Phi_k\left(\frac{\beta}{2},\vec{\theta}^k\right)} \equiv Z^{-1/2}e^{-(\beta/2)\hat{H}_0}\ket{\Phi_k}$. Similarly to Eq.~\eqref{Eq:expectation_value_in_energy_basis}, the expectation value of an observable $\hat{A}$ is then obtained by
\begin{equation}
\label{Eq:expectation_value_thermal_wavefunction}
\braket{\hat{A}}_\beta=\lim\limits_{K\rightarrow\infty}\frac{N}{K}\sum_{k=1}^{K}\Bra{\Phi\left(\frac{\beta}{2},\vec{\theta}^k\right)}\hat{U}^\dagger(t)\hat{A}\hat{U}(t)\Ket{\Phi\left(\frac{\beta}{2},\vec{\theta}^k\right)}\,.
\end{equation}
In practice, the limit in Eq.~\eqref{Eq:expectation_value_thermal_wavefunction} cannot be evaluated and some finite value of $K$ needs to be chosen. Evidently, if ${K>N}$, using an orthonormal basis and employing Eq.~\eqref{Eq:expectation_value_in_energy_basis} while propagating the full set of $N$ energy eigenstates is always preferable. For ${K<N}$ the random-phase sampling approach directly competes with truncated eigenstate sampling. 
In the following we show that sampling from eigenstates is the optimal approach to construct thermal wave functions, if no prior knowledge of the system dynamics is available and arbitrary observables shall be computed. We then compare this procedure to the random-phase approach in terms of its average behavior instead of the worst-case behavior in Sec.~\ref{subsec:ArbitraryObservables}. Before we begin, we first define an error measure to quantify how well a given sampling scheme performs regarding the approximation of time-dependent expectation values.

\section{Optimal sampling for arbitrary observables}
\label{sec:opt_sampling_arb_observable}

\subsection{Error measure}
\label{subsec:error_measure_arb_observable}
From this point onwards, $\hat{\rho}_\text{true}$ denotes the initial density operator as given by Eq.~\eqref{Eq:rho_true_in_energy_eigenbasis} and $\hat{\rho}_\text{approx}$ refers to any 
approximation of the density matrix, e.g., the truncated versions of Eq.~\eqref{Eq:rho_true_in_energy_eigenbasis} or Eq.~\eqref{Eq:rho_thermal}. We define the error of 
expectation values when using the approximated density matrix $\hat{\rho}_\text{approx}$ as
\begin{equation}
\varepsilon=\left|\text{tr}(\hat{A}\hat{U}(t)\hat{\rho}_\text{true}\hat{U}^\dagger(t))-\text{tr}(\hat{A}\hat{U}(t)\hat{\rho}_\text{approx}\hat{U}^\dagger(t))\right|\,.
\label{Eq:TrueError}
\end{equation}
In the following we are interested in finding an upper bound for this error. Since the error scales linearly with the norm of the observable $\hat{A}$, we consider only observables with a fixed Hilbert-Schmidt norm which we assume to be equal to one without loss of generality. In particular, the error of an arbitrary observable with unit Hilbert-Schmidt norm is bounded from above in the following way,
\begin{eqnarray}
\varepsilon&\leq&\max_{\norm{\hat{A}}_\text{HS}=1}\left|\text{tr}(\hat{A}\hat{U}(t)\hat{\rho}_\text{true}\hat{U}^\dagger(t))-\text{tr}(\hat{A}\hat{U}(t)\hat{\rho}_\text{approx}\hat{U}^\dagger(t))\right|\nonumber\\
	&=&\max_{\norm{\hat{A}}_\text{HS}=1}\left|\text{tr}(\hat{U}^\dagger(t)\hat{A}\hat{U}(t)\hat{\rho}_\text{error})\right|\,,\label{Eq:rho_error_def}
\end{eqnarray}
where we have used the abbreviation
\begin{equation}
\hat{\rho}_\text{error}=\hat{\rho}_\text{true}-\hat{\rho}_\text{approx}\,.
\end{equation}
For observables with arbitrary Hilbert-Schmidt norm, the error bound can simply be obtained by multiplication with the norm's value due to the error being a homogeneous function of $\hat{A}$. Equation \eqref{Eq:rho_error_def} can be reexpressed by evaluating the trace using the eigenbasis of $\hat{\rho}_\text{error}$ which we denote by $\{\ket{\xi_i}\}_{i=1,\dots,N}$,
\begin{eqnarray}
\varepsilon&\leq&\max_{\norm{\hat{A}}_\text{HS}=1}\left|\text{tr}(\hat{U}^\dagger(t)\hat{A}\hat{U}(t)\hat{\rho}_\text{error})\right|\nonumber\\
&=&\max_{\norm{\hat{A}}_\text{HS}=1}\left|\sum_{i=1}^{N}\bra{\xi_i}\hat{A}(t)\hat{\rho}_\text{error}\ket{\xi_i}\right|\nonumber\\
&=&\max_{\norm{\hat{A}}_\text{HS}=1}\left|\sum_{i=1}^{N}\bra{\xi_i}\hat{A}(t)\sum_{j=1}^{N}\ket{\xi_j}\bra{\xi_j}\hat{\rho}_\text{error}\ket{\xi_i}\right|\nonumber\\	
&=&\max_{\norm{\hat{A}}_\text{HS}=1}\left|\sum_{i=1}^{N}a_{ii}(t)\lambda_i\right|\,,\label{Eq:error_def_aii_lambdai}
\end{eqnarray}
where ${\hat{A}(t)=\hat{U}^\dagger(t)\hat{A}\hat{U}(t)}$ and $a_{ii}(t)=\bra{\xi_i}\hat{A}(t)\ket{\xi_i}$. Here we have used completeness, ${\hat{\mathbb{1}}=\sum_{j=1}^{N}\ket{\xi_j}\bra{\xi_j}}$, as well as the relation $\bra{\xi_j}\hat{\rho}_\text{error}\ket{\xi_i}=\lambda_i\delta_{ij}$, where $\lambda_i$ are the eigenvalues of $\hat{\rho}_\text{error}$. Note that, due to the unitary invariance of the Hilbert-Schmidt norm, the relation $\norm{\hat{A}}_\text{HS}=\norm{\hat{A}(t)}_\text{HS}$ holds for all times $t$. Then Eq.~\eqref{Eq:error_def_aii_lambdai} can be interpreted as a scalar product involving the vectorized quantities ${(\vec{a})_i=a_{ii}(t)}$ and ${(\vec{\lambda})_i=\lambda_i}$. Thus, the Cauchy-Schwarz inequality is applicable and it follows that
\begin{eqnarray}
\varepsilon&\leq&\max_{\lvert\vec{a}\rvert \leq 1}\left|\sum_{i=1}^{N}a_{ii}(t)\lambda_i\right|=\max_{\lvert\vec{a}\rvert \leq 1}|\vec{a}\cdot\vec{\lambda}|\nonumber\\
&\leq&\max_{\lvert\vec{a}\rvert \leq 1}|\vec{a}|\cdot|\vec{\lambda}|\nonumber=|\vec{\lambda}|\nonumber\\
&=&\norm{\hat{\rho}_\text{error}}_\text{HS}\,.\label{Eq:error_formula_derivation}
\end{eqnarray}
We have replaced ${\norm{\hat{A}}_\text{HS}=1}$ by ${\lvert\vec{a}\rvert \leq 1}$ since ${|\vec{a}|=\sqrt{\sum_i{a_{ii}^2}}\leq\sqrt{\sum_{i,j}{a_{ij}^2}}=\norm{\hat{A}}_\text{HS}}$. Note that if ${\hat{A}}$ and $\hat{\rho}_\text{error}$ are parallel, i.e., identical up to a scalar factor, the inequalities in Eq.~\eqref{Eq:error_formula_derivation} all become equalities. In conclusion, we obtain for the approximation error $\varepsilon$ the upper bound
\begin{equation}
\label{Eq:error_definition}
\varepsilon\leq\norm{\hat{\rho}_\text{error}}_\text{HS}=\sqrt{\sum_{i,j}\left[(\hat{\rho}_\text{error})_{ij}\right]^2}\,,
\end{equation}
which holds for arbitrary observables and arbitrary system dynamics - knowledge
of the initial state is entirely sufficient to evaluate the error bound.

\subsection{Optimal basis}
\label{subsec:optimal_basis_arb_observable}

The error bound in Eq.~\eqref{Eq:error_definition} is valid for any sampling method for $\hat{\rho}_\text{approx}$. This raises the question of the optimal sampling method. The following theorem yields the lowest attainable worst-case error bound for Eq.~\eqref{Eq:error_definition} as well as the corresponding sampling method which achieves this error:

\begin{theorem}
\label{prop:HS_Norm}
\normalfont{}
\textit{Let} ${\hat{\varrho}\in\mathbb{C}^{N\times N}}$ \textit{be an arbitrary Hermitian} ${N \times N}$ \textit{matrix. Then, for all} $\hat{M}\in\mathbb{C}^{N\times N}$ \textit{with} rank${(\hat{M}) \leq K}$\textit{, the inequality} ${\norm{\hat{\varrho}-\hat{M}}_\text{HS}^2\geq\sum_{i=K+1}^{N}|\lambda_i|^2}$ \textit{holds, where} ${\{\lambda_i\}_{i=K+1,\,\dots\,,N}}$ \textit{is the set containing the} $N-K$ \textit{smallest eigenvalues of} $\hat{\varrho}$\textit{. Equality is obtained if and only if} ${\hat{M}=\hat{P}\hat{\varrho}\hat{P}}$\textit{, where} $\hat{P}$ \textit{is a projector onto the eigenspace corresponding to the eigenvalue set} ${\{\lambda_i\}_{i=1,\,\dots\,,K}}$.
\end{theorem}
\textbf{Proof}
\normalfont{}
\begin{eqnarray}
\norm{\hat{\varrho}-\hat{M}}_\text{HS}^2&=&\langle \hat{\varrho}-\hat{M},\hat{\varrho}-\hat{M}\rangle_\text{HS}\nonumber\\
&=&\norm{\hat{\varrho}}_\text{HS}^2+\norm{\hat{M}}_\text{HS}^2-2\text{Re}\langle \hat{\varrho},\hat{M}\rangle_\text{HS}\,.\quad\label{Eq:HS_proof_equation_1}
\end{eqnarray} 
Since $\hat{M}$ has at most rank $K$ there exists a rank $K$ projector $\hat{P}$ such that ${\hat{M}=\hat{P}\hat{M} \hat{P}}$. It follows that
\begin{eqnarray}
\text{Re}\langle \hat{\varrho},\hat{M}\rangle_\text{HS}&\leq&\left|\langle \hat{\varrho},\hat{M}\rangle_\text{HS}\right|=\left|\tr{\hat{\varrho}\hat{P}\hat{M} \hat{P}}\right|\nonumber\\
&=&\left|\tr{\hat{P}\hat{\varrho}\hat{P}\hat{P}\hat{M} \hat{P}}\right|=\left|\langle \hat{P}\hat{\varrho}\hat{P},\hat{P}\hat{M} \hat{P}\rangle_\text{HS}\right|\nonumber\\
&\leq&\norm{\hat{P}\hat{\varrho}\hat{P}}_\text{HS}~\norm{\hat{P}\hat{M} \hat{P}}_\text{HS}\,.\label{Eq:HS_proof_PEP_HS_PPsiP_HS}
\end{eqnarray}
Here, the projector property $\hat{P}=\hat{P}^2$, invariance under cyclic permutation of the trace, and the Cauchy-Schwarz inequality have been used. Minimizing Eq.~\eqref{Eq:HS_proof_equation_1} translates into maximizing Eq.~\eqref{Eq:HS_proof_PEP_HS_PPsiP_HS}, where the maximum, i.e.~equality, is given only if ${\hat{P}\hat{M} \hat{P}\parallel \hat{P}\hat{\varrho}\hat{P}\Leftrightarrow \hat{P}\hat{M} \hat{P}=\mu \hat{P}\hat{\varrho}\hat{P}}$ with $\mu\in\mathbb{C}$. Inserting this relation into Eq.~\eqref{Eq:HS_proof_PEP_HS_PPsiP_HS}, we obtain $\text{Re}(\mu)=|\mu|$ which implies $\mu\in\mathbb{R}_0^+$. In the following we focus on determining $\hat{P}$ such that $\mu\norm{\hat{P}\hat{\varrho}\hat{P}}_\text{HS}^2$ is maximal. This is obtained by plugging in this parallelity condition into the final expression in Eq.~\eqref{Eq:HS_proof_PEP_HS_PPsiP_HS}. 

Representing $\hat{\varrho}$ and $\hat{P}$ in a basis $\{\ket{i}\}_{i=1,\dots,N}$ which diagonalizes $\hat{P}$, i.e.,
\begin{equation}
\hat{P}=\sum_{i=1}^{K}\ket{i}\bra{i}\qquad\text{and}\qquad \hat{\varrho}=\sum_{i,j=1}^{N}\varrho_{ij}\ket{i}\bra{j}\,,
\end{equation}
it follows that
\begin{eqnarray}
\mu\norm{\hat{P}\hat{\varrho}\hat{P}}_\text{HS}^2&=&\mu\,\tr{\hat{P}\hat{\varrho}\hat{P}\hat{P}\hat{\varrho}\hat{P}}\nonumber\\
&=&\mu\,\tr{\sum_{i,j,k=1}^{K}\varrho_{ij}\varrho_{jk}\ket{i}\bra{k}}\nonumber\\
&=&\mu\sum_{i,j=1}^{K}\varrho_{ij}\varrho_{ji}=\mu\sum_{i,j=1}^{K}\varrho_{ij}\varrho^{*}_{ij}\nonumber\\
&=&\mu\sum_{i,j=1}^{K}|\varrho_{ij}|^2\nonumber=\mu\sum_{i,j=1}^{K}\lvert\bra{i}\hat{\varrho}\ket{j}\rvert^2\,.\label{Eq:HS_proof_iEj}
\end{eqnarray}
Representing $\hat{\varrho}$ in its eigenbasis $\{\ket{\phi_k}\}_{k=1,\dots,N}$, i.e.,
\begin{equation}
\hat{\varrho}=\sum_{k=1}^{N}\lambda_{k}\ket{\phi_k}\bra{\phi_k}\,,\label{Eq:HS_proof_E_def_eigenbasis}
\end{equation} 
with eigenvalues $\{\lambda_k\}_{k=1,\dots,N}$ sorted such that ${\lambda_1\geq\dots\geq \lambda_N}$, Eq.~\eqref{Eq:HS_proof_iEj} translates into
\begin{eqnarray}
\mu\sum_{i,j=1}^{K}\lvert\bra{i}\hat{\varrho}\ket{j}\rvert^2&=&\mu\sum_{i,j=1}^{K}\sum_{k=1}^{N}|\lambda_k|^2\,\lvert\braket{i|\phi_k}\rvert^2\,\lvert\braket{j|\phi_k}\rvert^2\nonumber\\
&=&\mu\sum_{k=1}^{N}|\lambda_k|^2\,\left[\sum_{i=1}^{K}\lvert\braket{i|\phi_k}\rvert^2\right]^2\,.\label{Eq:e_k_iE_k}
\end{eqnarray}
Now we define $z_k=\sum_{i=1}^{K}\lvert\braket{i|\phi_k}\rvert^2$. The inequality ${z_k\leq 1}$ holds since ${\sum_{i=1}^{K}\lvert\braket{i|\phi_k}\rvert^2\leq\sum_{i=1}^{N}\lvert\braket{i|\phi_k}\rvert^2=1}$ applies. It directly follows that ${z_k^2\leq z_k\leq 1}$ with ${z_k^2=z_k=1}$ if and only if $\ket{\phi_k}\in\text{im}(\hat{P})$, with $\text{im}(\hat{P})$ the image of $\hat{P}$. Furthermore ${\sum_{k=1}^{N}z_k^2\leq K}$ holds with 
\begin{equation}
{\sum_{k=1}^{N}z_k^2=K} \Leftrightarrow \forall\,k\text{:} \ket{\phi_k}\in\text{im}(\hat{P})\,,\label{Eq:z_k_squared_equal_K}
\end{equation}
which can be seen as follows,
\begin{eqnarray}
\sum_{k=1}^{N}z_k^2&\leq&\sum_{k=1}^{N}z_k=\sum_{k=1}^{N}\sum_{i=1}^{K}\braket{i|\phi_k}\braket{\phi_k|i}\nonumber\\
&=&\sum_{i=1}^{K}\braket{i|i}=K\,.
\end{eqnarray}
Equations \eqref{Eq:HS_proof_iEj} and \eqref{Eq:e_k_iE_k} imply, that
\begin{equation}
\lambda\norm{\hat{P}\hat{\varrho}\hat{P}}_\text{HS}^2=\mu{•}{•}\sum_{k=1}^{N}|\lambda_k|^2\,z_k^2\leq\mu\sum_{k=1}^{K}|\lambda_k|^2\,,
\end{equation}
with equality if $z_k=1$ when $k$ is an index belonging to a set containing the $K$ largest eigenvalues of $\hat{\varrho}$ and $z_k=0$ otherwise. Thus $\text{im}(\hat{P})$ needs to be spanned by a set of eigenvectors corresponding to such an eigenvalue set.

Now we can finally rewrite Eq.~\eqref{Eq:HS_proof_equation_1},
\begin{eqnarray}
\norm{\hat{\varrho}-\hat{M}}_\text{HS}^2&=&\norm{\hat{\varrho}}_\text{HS}^2+\norm{\hat{P}\hat{M} \hat{P}}_\text{HS}^2-2\text{Re}\langle \hat{\varrho},\hat{M}\rangle_\text{HS}\nonumber\\
&=&\norm{\hat{\varrho}}_\text{HS}^2+\mu^2\norm{\hat{P}\hat{\varrho}\hat{P}}_\text{HS}^2-2\text{Re}\langle \hat{\varrho},\hat{M}\rangle_\text{HS}\nonumber\\
&{\geq}&\norm{\hat{\varrho}}_\text{HS}^2+\mu^2\norm{\hat{P}\hat{\varrho}\hat{P}}_\text{HS}^2-2\mu\norm{\hat{P}\hat{\varrho}\hat{P}}_\text{HS}^2\,,\nonumber\label{Eq:minimize_E_minus_psi}\\
\end{eqnarray}
where we have used Eq.~\eqref{Eq:HS_proof_PEP_HS_PPsiP_HS} in the final line. Searching for the extremum of this expression with respect to $\mu$,
\begin{equation}
0\overset{!}{=}\frac{\partial}{\partial\mu} \norm{\hat{\varrho}-\hat{M}}_\text{HS}^2=(2\mu-2)\norm{\hat{P}\hat{\varrho}\hat{P}}_\text{HS}^2\,,
\end{equation}
we find that $\mu=1$ uniquely minimizes Eq.~\eqref{Eq:minimize_E_minus_psi}. 
Consequently,
\begin{equation}
\norm{\hat{\varrho}-\hat{M}}_\text{HS}^2\geq\sum_{k=K+1}^{N}|\lambda_k|^2\,
\end{equation}
holds with equality if and only if ${\hat{M}=\hat{P}\hat{\varrho}\hat{P}}$ with $\text{im}(\hat{P})$ being spanned by the eigenvectors corresponding to the $K$ largest eigenvalues of $\hat{\varrho}$. This concludes the proof.
\hfill $\square$\\

Theorem~\ref{prop:HS_Norm} can be applied to sampling mixed states via a set of pure states in the following way: Choose ${\hat{\varrho}=\hat{\rho}_\text{true}}$ and ${\hat{M}=\hat{\rho}_\text{approx}}$. Then constructing the density matrix ${\hat{\rho}_\text{approx}}$ in the eigenbasis of ${\hat{\rho}_\text{true}}$,
\begin{equation}
(\hat{\rho}_\beta)_{ij}=\begin{cases} \frac{1}{Z}e^{-\beta E_i} &\text{if}~i=j~\mathrm{and}~i<K\,, \\
0 & \text{otherwise}\,, \end{cases}
\end{equation}
minimizes the worst case for the error $\varepsilon$, cf.~Eq.~\eqref{Eq:rho_error_def}, uniquely.
The first $K$ diagonal entries of ${\hat{\rho}_\text{approx}}$ are the largest eigenvalues of ${\hat{\rho}_\text{true}}$, hence the worst-case error can be estimated via
\begin{eqnarray}
\varepsilon&\leq&\norm{\hat{\rho}_\text{true}-\hat{\rho}_\text{approx}}_\text{HS}=\sqrt{\sum_{i=K+1}^{N}\lvert \lambda_i\rvert^2}\nonumber\\
&=&\sqrt{\sum_{i=K+1}^{N}p_i^2}\equiv\varepsilon_\text{bound}(K) \,. \label{Eq:sum_lowest_eigenvalues}
\end{eqnarray}
This error bound only depends on the eigenvalues $\lambda_i$ of ${\hat{\rho}_\text{true}}$. It is particularly small if the initial density matrix possesses a narrow population distribution. For thermal systems, this is usually connected to low temperatures. Furthermore, the error bound is independent of the actual system dynamics. As a result, if no prior information is available and one aims to minimize the worst-case error for computing observables of an ensemble, propagating the eigenvectors of the initial density matrix corresponding to the largest eigenvalues is the uniquely optimal choice.

\section{Accelerating convergence}
\label{sec:sampling_enhancements}

In this section we will discuss two enhancements to the random-phase wave function approach and eigenstate-based sampling which substantially accelerate their convergence in certain regimes. 
The fundamental idea underlying these improvements is to eliminate redundant information from the sampling procedures. Definining the traceless part of the density matrix, $\hat{\rho}_0=\hat{\rho}-\hat{\mathbb{1}}\cdot\text{tr}[\hat{\rho}]$, we can expand the evolution of the density matrix in the Schr{\"o}dinger picture as follows,
\begin{eqnarray}
\hat{\rho}(t)=\hat{U}^\dagger\hat{\rho}(0)\hat{U}&=&\hat{U}^\dagger\left[\hat{\mathbb{1}}\cdot\text{tr}[\hat{\rho}]+\hat{\rho}_0(0)\right]\hat{U}\nonumber\\&=&\hat{\mathbb{1}}\cdot\text{tr}[\hat{\rho}]+\hat{U}^\dagger\hat{\rho}_0(0)\hat{U}\,.\label{Eq:schroedinger_picture}
\end{eqnarray}
A similar expansion in the Heisenberg picture leads to the following result,
\begin{eqnarray}
\hat{A}(t)=\hat{U}^\dagger\hat{A}(0)\hat{U}&=&\hat{U}^\dagger\left[\hat{\mathbb{1}}\cdot\text{tr}[\hat{A}]+\hat{A}_0(0)\right]\hat{U}\nonumber\\&=&\hat{\mathbb{1}}\cdot\text{tr}[\hat{A}]+\hat{U}^\dagger\hat{A}_0(0)\hat{U},\label{Eq:heisenberg_picture}
\end{eqnarray}
with $\hat{A}_0=\hat{A}-\hat{\mathbb{1}}\cdot\text{tr}[\hat{A}]$. 
As a consequence Eq.~\eqref{Eq:schroedinger_picture} and Eq.~\eqref{Eq:heisenberg_picture} show that the identity component can be isolated on both operator and state level. Most notably, this contribution
to the state, respectively the observable, is independent of the actual propagation.
It is only the traceless part of the state, respectively the observable, which is affected by the dynamics and thus requires sampling. In the following, we
investigate how the
redundant information corresponding to the identity component 
is treated by the sampling protocols.
This allows us to formulate improved sampling schemes which systematically discard the identity components.

\subsection{Shifting the spectrum of the observable}
\label{subsec:shifting_the_spectrum_of_the_observable}

Shifting an observable $\hat{A}$ by a constant $\lambda \hat{\mathbb{1}}$
merely shifts the expectation value $\braket{\hat{A}}$ by $\lambda$ due to 
linearity of the expectation value and normalization of the states,
cf.~Eq.~\eqref{Eq:expectation_value_in_energy_basis}.
Since for random-phase sampling the diagonal entries of the initial density
matrix are faithfully represented, 
expectation values are not affected by homogeneous spectral shifts. 
In particular, the error incurred when 
approximating the expectation value of an observable $\hat{A}$ and the spectrally shifted observable $\hat{A} + \lambda \hat{\mathbb{1}}$ is identical, i.e.~$\varepsilon(\braket{\hat{A}})=\varepsilon(\braket{\hat{A}+\lambda \hat{\mathbb{1}}})\,\forall\lambda$. However, this invariance does not hold for eigenstate-based sampling,
\begin{eqnarray}
\varepsilon(\braket{\hat{A}+\lambda \hat{\mathbb{1}}})&=&\left\vert\sum_{n=K+1}^{N}p_n\bra{E_n}\hat{U}^\dagger(\hat{A}+\lambda \hat{\mathbb{1}})\hat{U}\ket{E_n}\right\vert\nonumber\\
&=&\left\vert\underbrace{\sum_{n=K+1}^{N}p_n\bra{E_n}\hat{U}^\dagger\hat{A}\hat{U}\ket{E_n}}_{\varepsilon_1}+\underbrace{\lambda\hspace*{-0.1cm}\sum_{n=K+1}^{N}p_n}_{\varepsilon_2}\right\vert\nonumber\\
&\neq& \left\vert \varepsilon_1 \right\vert = \varepsilon(\braket{\hat{A}})\,.
\end{eqnarray}
Specifically, if the error contribution $\varepsilon_2$ is much larger than the error $\varepsilon_1$, then the overall error will be dominated by a component which is entirely caused by the non-vanishing trace of the observable. As pointed out in the beginning of this section such a component carries only redundant information and thus can and should be removed from the sampling procedure.

To this end, we modify the eigenbasis sampling presented in Eq.~\eqref{Eq:eigenbasis_sampling} by separating the observable $\hat{A}$ into its traceless part, $\hat{A}_0$, and a multiple of identity, $\lambda_0\hat{\mathbb{1}}=\text{tr}[\hat{A}]\hat{\mathbb{1}}$, such that ${\hat{A}=\hat{A}_0+\lambda_0\hat{\mathbb{1}}}$. This allows to reexpress the approximated expectation value as follows,
\begin{eqnarray}
\braket{\hat{A}}&\approx&\sum_{n=1}^{K}p_n\bra{E_{n}}\hat{U}^\dagger(t)(\hat{A_0}+\lambda_0\hat{\mathbb{1}})\hat{U}(t)\ket{E_{n}}\nonumber\\
&=&\underbrace{\sum_{n=1}^{K}p_n\bra{E_{n}}\hat{U}^\dagger(t)\hat{A}_0\hat{U}(t)\ket{E_{n}}}_{\stackrel{K\rightarrow N}{=}\braket{\hat{A}_0}}+\underbrace{\lambda_0\sum_{n=1}^{K}p_n}_{\stackrel{K\rightarrow N}{=}\text{tr}[\hat{A}]}\label{Eq:EigenbasisShift}\\
&\approx&\sum_{n=1}^{K}p_n\bra{E_{n}}\hat{U}^\dagger(t)\hat{A}_0\hat{U}(t)\ket{E_{n}}+\lambda_0\label{Eq:EigenbasisShiftFix}\,.
\end{eqnarray}
Note, that in the limit $K \rightarrow N$ the second contribution in Eq.~\eqref{Eq:EigenbasisShift} is simply given by $\text{tr}[\hat{A}]$, which is a property that is known \textit{a priori} since it neither depends on the initial ensemble nor on the dynamics. Thus, a sampling of this quantity would be redundant and could artificially inflate the error of the protocol. Fortunately, this issue can be easily fixed in Eq.~\eqref{Eq:EigenbasisShiftFix} by replacing the observable with its traceless version and manually adding the contribution from the trace in the very end. As a side effect this adjustment also ensures that, just like the random-phase approach, eigenstate-based sampling becomes invariant with respect to homogeneous spectral shifts.

\subsection{Removing the background}
\label{subsec:removing_the_background}

The issue of redundant information due to the contribution of identity arises not only for the observables, but also for the initial ensemble.
To elucidate this fact, we define the
reduced populations $p'_n$ and the minimal population $p_\text{min}$ as follows,
\begin{equation}
p'_n=p_n-p_\text{min}\,,\qquad p_\text{min}=\text{min}(p_n)\,.
\end{equation}
This allows to separate the identity component of a density matrix, $p_\text{min}\cdot\hat{\mathbb{1}}$, from its positive remainder\footnote{Note that, in order to preserve
positivity, we do not set the trace of the adjusted density matrix to zero. To reflect upon this change, we distinguish the reduced density matrix $\hat{\rho}'$ from the traceless density matrix $\hat{\rho}_0$. It is an open question whether additional improvements can be obtained by foregoing positivity which we leave up to future work.} $\hat{\rho}'$, 
\begin{equation}
\hat{\rho}=\hat{\rho}'+p_\text{min}\cdot\hat{\mathbb{1}}\,.
\end{equation}
Since the identity component $p_\text{min}\cdot\hat{\mathbb{1}}$ is 
unaffected by time evolution, it can be considered a redundant background and 
should be removed from the dynamics beforehand. This becomes particularly clear 
when we express expectation values in terms of the reduced populations and 
additionally separate the identity component of the observable as explained above, 
\begin{eqnarray}
\braket{\hat{A}}
&=&\sum_{n=1}^{N}(p'_n+p_\text{min})\bra{E_{n}}\hat{U}^\dagger(t)\hat{A}_0\hat{U}(t)\ket{E_{n}}+\lambda_0\nonumber\\
&=&\sum_{n=1}^{N}p'_n\bra{E_{n}}\hat{U}^\dagger(t)\hat{A}_0\hat{U}(t)\ket{E_{n}}+\lambda_0+p_\text{min}\underbrace{\text{tr}[\hat{A}_0]}_{=0}\nonumber\\
&=&\sum_{n=1}^{N}p'_n\bra{E_{n}}\hat{U}^\dagger(t)\hat{A}_0\hat{U}(t)\ket{E_{n}}+\lambda_0\,.\label{Eq:ReducedPopulationsEigenbasis}
\end{eqnarray}

\begin{figure}[t]
	\centering	\includegraphics[width=1.0\columnwidth]{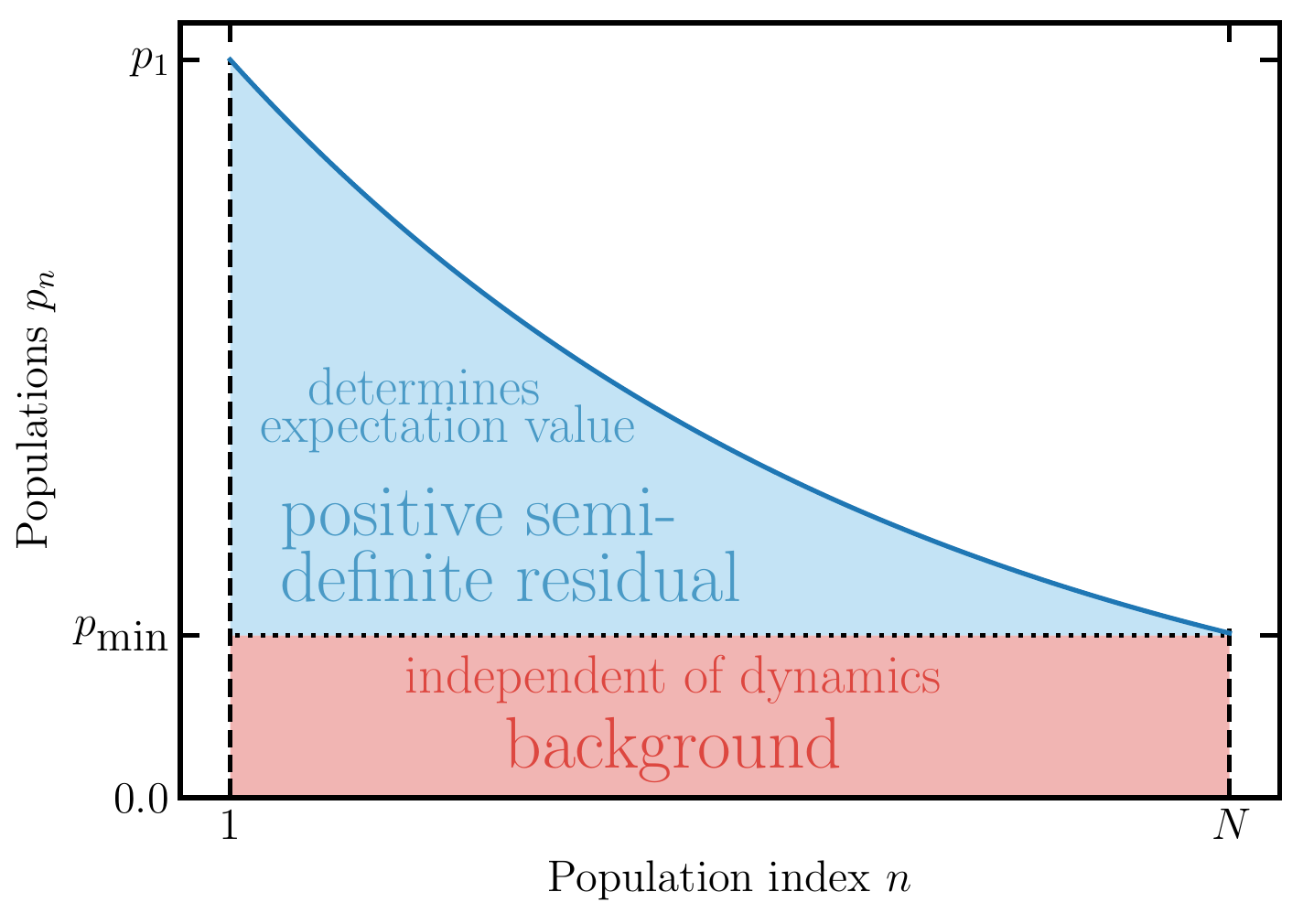}
	\caption{Population as a function of energy in a typical Boltzmann distribution. The horizontal dotted line at $p_N$ distinguishes those parts of the population for individual states which can change due to unitary time evolution (blue) compared to the ``background'' which always remains unchanged (red).}
	\label{Fig:PopulationSketch}
\end{figure}

A visualization of this ``background removal'' is shown in Fig.~\ref{Fig:PopulationSketch} for the example of a thermal ensemble. 
The background (red) is proportional to the minimal population $p_\text{min}$. 
Only the contribution due to the reduced coefficients $p_n'$ (blue) is affected by the dynamics and thus needs to be sampled, cf. Eq.~\eqref{Eq:ReducedPopulationsEigenbasis}. Note that Theorem~\ref{prop:HS_Norm} still holds when sampling a density matrix from which the minimal population component is
subtracted, i.e., 
\begin{equation}
\varepsilon_\text{bound}(K)=\sqrt{\sum_{i=K+1}^{N}{p_i'}^2}\,.
\label{Eq:BoundReduced}
\end{equation}
This expression yields the worst-case error bound for the eigenstate-based sampling scheme with
the two improvements discussed above. It is particularly evident how the
removal of the minimal population leads to decreased values for the reduced
populations $p_i' = p_i - p_{\text{min}} \leq p_i$ and thus
also to a reduced approximation error.

Note that the removal of the background can also be employed for the random-phase 
approach by constructing the random-phase thermal wave functions using the 
reduced populations,
\begin{equation}
\Ket{\Phi_k\left(\frac{\beta}{2},\vec{\theta}^k\right)}=\frac{1}{\sqrt{N}}\sum_{j=1}^{N}\sqrt{p_j'}e^{i\theta_j^k}\ket{E_j}\,.
\end{equation}
For traceless observables, using these adjusted wave functions
still ensures convergence towards the exact expectation value for
$N\rightarrow \infty$, following an argument analogous to the one for the  eigenstate-based approach presented
above.

\section{Application to a spin chain}
\label{sec:app_to_random_observables}

\subsection{General Observables}
\label{subsec:ArbitraryObservables}

Theorem~\ref{prop:HS_Norm} implies
that randomized sampling using random-phase thermal wave functions yields an inferior worst-case error bound compared to deterministic sampling using the lowest-lying eigenstates. However, it does not allow us to make a statement on whether this relationship holds in a typical setting, i.e., \textit{on average}, since the average error might behave rather differently from the worst-case bound. In order to investigate this point we compare random-phase sampling of thermal wave functions with eigenstate-based sampling using the averaged results of randomized simulations. Moreover we
also compare the performance of these two methods when employing the improvements discussed in Sec.~\ref{sec:sampling_enhancements}.

To gain insight into the scaling of the average error with Hilbert space dimension in a typical physical setting, we study a one-dimensional thermal spin chain with nearest-neighbor coupling, see, e.g., Refs.~\cite{Brennen2004,Maziero2010,Novotny2016}. The corresponding Hamiltonian reads
\begin{equation}
\hat{H}(t)=-J\sum_{j=1}^{n-1}\hat{\vec\sigma}_j\hat{\vec\sigma}_{j+1}-h_z\sum_{j=1}^{n}\hat{\sigma}_j^z+\mathcal{E}(t)\sum_{j=1}^{n}\hat{\sigma}_j^x\,,
\label{Eq:HamiltonianSpinChain}
\end{equation}
with $J$ the coupling strength, $n$ the number of spins, $h_z$ the field strength of an external magnetic field in $z$-direction, and $\mathcal{E}(t)$ representing a time-dependent magnetic field in $x$-direction with a truncated Gaussian envelope such that $\mathcal{E}(t=0)=0$. The vector operator $\hat{\vec\sigma}_j=(\hat{\sigma}_j^x,\hat{\sigma}_j^y,\hat{\sigma}_j^z)$ contains the Pauli matrices acting on spin $j$. Increasing the length of the spin chain doubles the Hilbert space dimension, exemplifying the curse of dimensionality present in almost all quantum systems. 

To address the question on how to properly simulate the average error in practical applications we first motivate why a straightforward approach of completely averaging over all possible observables and dynamics is ill-advised. Such a complete averaging can be performed by using Haar-measure randomly distributed unitaries~\cite{MezzadriAMS2007} for both the unitary propagator and the observable eigenbases. We performed reference calculations which implemented such a full averaging procedure. We observed that the corresponding results significantly deviate from simulation results we obtain when only using limited resources, however, we argue that the former averaging process is not physically sensible. Using Haar-measure random unitaries indiscriminately mixes all regions in Hilbert space, which leads to a high degree of averaging such that barely any structure remains. More specifically we observed that in combination with the sampling improvements outlined in Sec.~\ref{sec:sampling_enhancements} almost all expectation values become very close to zero - an effect that is further amplified when the Hilbert space dimension increases. Since the simulation of physical observables is aimed at probing certain structures in physical process, we therefore think that using an approach based on uniform drawing according to the Haar measure will not yield a fair assessment on the behaviour of the sampling algorithms in practical applications.

For this reason our results on the average performance were not obtained using a complete mathematical average but rather in the following way: First, a random observable is constructed by building randomized linear combinations of single- and two-particle observables. Secondly, different time evolutions are sampled by constructing driving fields $\mathcal{E}(t)$ formed by a Gaussian envelope using a set of randomly drawn carrier frequencies with limited bandwith. Accounting for the discretization of time in our numerical setup, the field $\mathcal{E}(t)$ at times $t_n$ with $0\leq n<N_T$ is given by
\begin{equation}
\mathcal{E}(t_n)\sim \mathcal{E}_0e^{-{\frac{(t_n-t')^2}{2\tau^2}}}\cdot \sum_{j=0}^{N_T}e^{-2\pi i \frac{j(t_n-t')}{N_T}}\bar{a}_j\,,
\end{equation}
where $N_T$ represents the number of discrete timesteps, $\mathcal{E}_0$ scales the field strength, $t'$ is a constant shift in time, and $\tau$ is the width of the Gaussian envelope. With this construction the temporal profile is generated from randomly drawn frequency amplitudes $\bar{a}_j$ by discrete Fourier transformation. In order to obtain physically sensible pulses with a proper numerical representation, only the smallest 5\% of the frequencies that can be realized on our time grid are allowed to be non-zero.

Our choice to restrict observables to linear combinations of single- and two-particle operators is motivated by the fact that many observables of physical interest are local. Furthermore the bandwith restriction we employed on our driving fields takes into account that in practical applications there are usually physical limitations on energy and time resolution. Our simulation parameters in atomic units are $J=1$, $h_z=0.002$ and the envelope of the driving field $\mathcal{E}(t)$ has a width of $\tau=170$. $\mathcal{E}_0$ is chosen such that the pulse has a maximal peak amplitude of $\mathcal{E}_\text{max}=1.0$. The initial state of the system is chosen as a thermal canonical ensemble with temperature T, which uniquely corresponds to a purity $P$. 

A subtle issue is encountered when normalizing all observables with respect to the Hilbert-Schmidt norm as discussed in Sec.~\ref{subsec:error_measure_arb_observable}. Using observables with unit Hilbert-Schmidt norm leads to expectation values at final time that become ever closer to zero when increasing the length of the spin chain, i.e., the Hilbert space dimension. If left unaccounted for, this effect would severely impede our ability to analyze the behavior of the average absolute error given by Eq.~\eqref{Eq:TrueError} as a function of system size. To ensure comparability of average errors obtained for arbitrary Hilbert space dimensions we thus aimed to enforce in our simulations that the standard deviation of our randomly generated observables' spectra is constant. This translates to a dimensionally independent spread of expectation values. Note that in our simulations the mean expectation value is zero due to the tracelessness of our observables, cf.~Sec.~\ref{subsec:shifting_the_spectrum_of_the_observable}. This explains our observation of decreasing expectation values when using normalized observables as discussed in Sec.~\ref{subsec:error_measure_arb_observable}. It can be shown that in order to achieve such a constant spread one needs to rescale all randomly generated observables to a Hilbert-Schmidt norm of $\sqrt{N}$ instead of one. Note that this scaling of the observables rescales all expectation values by the same amount but leaves the sampling approaches otherwise unaffected. In particular it does not change the relative performance of the two methods in any way. 

We also want to point out that the sampling enhancements we introduce in Sec.~\ref{sec:sampling_enhancements} alter the Hilbert-Schmidt norm of both observable and initial density matrix. However, it is important to note that we do not rescale these quantities uniformly. Our enhancements achieve a reduction of the Hilbert-Schmidt norm in such a way that we can still obtain an estimate for the original expectation values involving the original Hilbert-Schmidt norm for both observable and density matrix. In particular this goes beyond simply multiplying the observable with a prefactor smaller than one to obtain a smaller error. After all, such a manipulation would also decrease the expectation values we aim to approximate. Even though a reduction of the absolute errors could be achieved in that manner, the relative errors would remain unchanged. In contrast, our enhancements reduce both the absolute error as well as the relative error of our estimates.

Finally, we decided in this work to consistently use the energy eigenbasis to construct the random-phase wave functions in order to obtain a fair comparison with the eigenstate-based approach. Specifically, this choice removes any effects originating from a different basis choice between the two approaches which might obscure differences in convergence behaviour that are inherent to the two methods. Note that in large systems, performing a full diagonalization to obtain the eigenstates of the initial ensemble may pose an issue. For the construction of random-phase thermal wave functions it is possible to use an arbitrary orthonormal basis instead, cf.~Sec.~\ref{subsec:thermal_sampling}. For the eigenstate-based approach there is unfortunately no way around calculating the corresponding eigenstates. However, in a practical application one should expect the required amount of sampling states for a reasonable sampling error to be much smaller than the total Hilbert space dimension. In this case algorithms to determine a small amount of eigenstates, e.g.~the Arnoldi method~\cite{Arnoldi1951}, can be employed.

\begin{figure}[t]
	\centering
	\includegraphics[width=1.0\columnwidth]{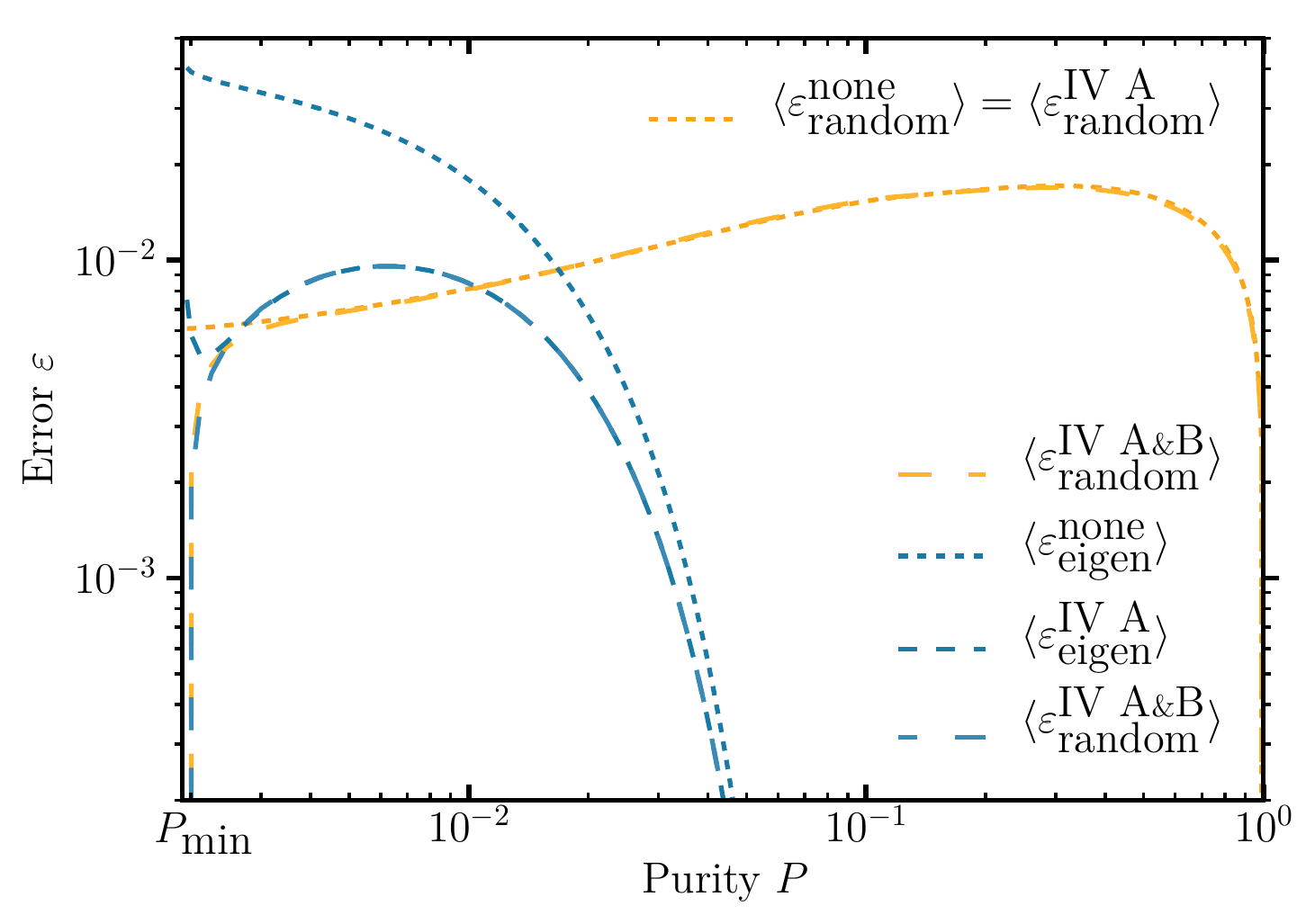}
  \caption{Average error for expectation values obtained for 150 randomized observables in the eigenstate-based approach (orange) and using random-phase sampling (blue) for a spin chain with 9 spins using $K=10$ states in the sampling. Short dashed lines refer to sampling schemes with no adjustments, whereas medium dashed lines include the improvement discussed in Sec.~\ref{subsec:shifting_the_spectrum_of_the_observable}. On top of this, long dashed lines also show the performance when the removal of the background in the initial ensemble has been additionally included, cf.~Sec.~\ref{subsec:removing_the_background}. The average error for the random-phase sampling with respect to no sampling adjustments and to sampling adjustments following Sec.~\ref{subsec:shifting_the_spectrum_of_the_observable} coincide and are just represented by a single short dashed line.}
	\label{Fig:ErrorImprovementComparison}
\end{figure}

We begin by analyzing the behavior of the sampling algorithms when employing the enhancements proposed in Sec.~\ref{sec:sampling_enhancements}. Figure~\ref{Fig:ErrorImprovementComparison} compares the average absolute error, cf.~Eq.~\eqref{Eq:TrueError}, obtained with eigenstate-based sampling (orange), respectively random-phase sampling (blue), once without any enhancements (short dashed lines) and once with removal of the observable trace only (medium dashed lines). In addition to that, the performance when using both improvements as discussed in Sec.~\ref{sec:sampling_enhancements} is also shown (long dashed lines).
The data for Fig.~\ref{Fig:ErrorImprovementComparison}
was obtained for 9 spins and a sampling size of
$K=10$ eigenstates, respectively $K=10$ random-phase wave functions. 
The absolute error in the expectation value was averaged over a set of 150 
random observables. We performed our simulations using an initial thermal 
ensemble for a wide range of different temperatures. To improve comparability
for different physical systems, we decided to plot this error against the purity
of the ensemble instead of the temperature. Note that the purity is a constant
of motion in the coherent dynamics we are simulating.

Not surprisingly, we observe that eigenstate-based sampling performs 
particularly well for high purity, i.e.~low temperatures, whereas the 
random-phase approach performs about equally well for most purities. The error
drop towards purities of $P=1$ is due to the fact that, for a thermal ensemble
that is close to zero temperature, all random-phase wave function become
similar to the ground state of the system. As a consequence, in the limit of purity
$P=1$ (or temperature $T=0$), only a single realization is required to obtain 
exact results for the expectation values. Naturally, a similar argument can be
made on the eigenstate-based approach, where the sampling is progressively
performed starting from the energetically lowest-lying eigenstates. However, we
observe that the performance of the eigenstate-based sampling remains stable for
a much larger purity window around $P=1$ compared to the random-phase approach.

When the ensemble purity decreases, i.e.~the temperature increases, the 
performance of the eigenstate-approach quickly deteriorates and for purities
around $10^{-2}$ it becomes comparable to the random-phase sampling. At these 
temperatures the adjustments we introduced above become relevant. With respect to
eigenstate-based sampling it is evident that removing the trace of the observables we
discussed in Sec.~\ref{subsec:shifting_the_spectrum_of_the_observable} leads to
a clear performance boost of the sampling scheme.
The short dashed lines in Fig.~\ref{Fig:ErrorImprovementComparison} represents 
the average error of expectation values obtained for non-traceless 
randomized observables. In order to generate non-traceless random observables, 
we add a multiple of identity to traceless random observables, with the 
corresponding prefactor 
chosen in such a way that it has the same order of 
magnitude as an average element of the original observable. This ensures that
the error we introduce by homogeneously shifting the spectrum of the observable
is not arbitrarily large but remains within reasonable bounds with respect to
physical setups. 

Comparing the 
original (short dashed blue line) to the adjusted (medium dashed blue line) 
eigenstate-based sampling we observe a performance improvement of up to one order of
magnitude at minimal purity. At high purities this correction becomes less relevant.
Yet, in the vicinity of minimal purity, i.e.~if the initial ensemble is close 
to the maximally mixed state, the removal of the background has by far the strongest 
effect. Without this adjustment to the sampling procedure, both eigenstate-based sampling and the 
random-phase approach show a non-zero error at minimal purity.
Conversely, when background removal is employed (long dashed lines), the performance
of both methods can be dramatically increased in the neighborhood of the maximally mixed
state. Most notably, at minimal purity, the error for both methods goes exactly
to zero which properly accounts for the fact, that the maximally mixed state
contains no structure and as such all observable
expectation values are known \textit{a priori} (they are equal to the
observable's trace) and do not need to be sampled at all. Due to the clear 
success of our improvements in accelerating convergence, we will in the 
following show exclusively the results we obtained when using the two sampling
methods with both enhancements introduced in
Sec.~\ref{sec:sampling_enhancements}.

\begin{figure*}[tb]
	\centering	\includegraphics[width=2.0\columnwidth]{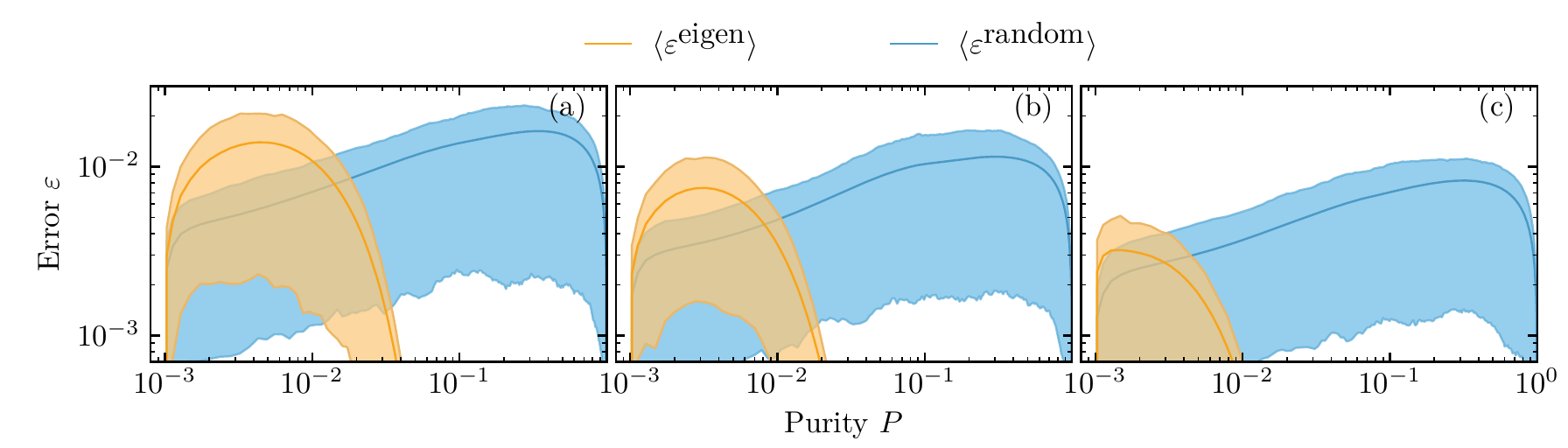}
  \caption{Average error for expectation values obtained for 200 randomized observables in the eigenstate-based approach (orange) and using random-phase sampling (blue) for a spin chain with 10 spins (Hilbert space dimension $N=2^{10}=1024$). Results were obtained using $K=0.05 N$ (a), $K=0.10 N$, and $K=0.20 N$ states in the samples. The shaded regions show the error range inside which 80\% of all performed simulations fall.}
	\label{Fig:AccuracyComparison}
\end{figure*}

Using the enhanced sampling algorithms, we begin by turning our study towards 
the sampling methods' performance with respect to the number of sampling 
elements $K$. The solid lines in Fig.~\ref{Fig:AccuracyComparison} show the 
average error for expectation values in a spin chain with 10 spins for $K$
corresponding to 5\%, 10\%, and 20\% of the Hilbert space size dimension $N$.
The shaded areas around these lines show the error region inside which 80\% of
our simulations with randomly drawn observables and time evolutions fall.
We chose this representation to illustrate the variation in our results instead 
of a simple standard deviation to account for the fact that the errors we obtain
scatter quite asymmetrically around the mean, particularly in the regions near
minimal and maximal purity.
\begin{figure*}[tb]
	\centering	\includegraphics[width=2.0\columnwidth]{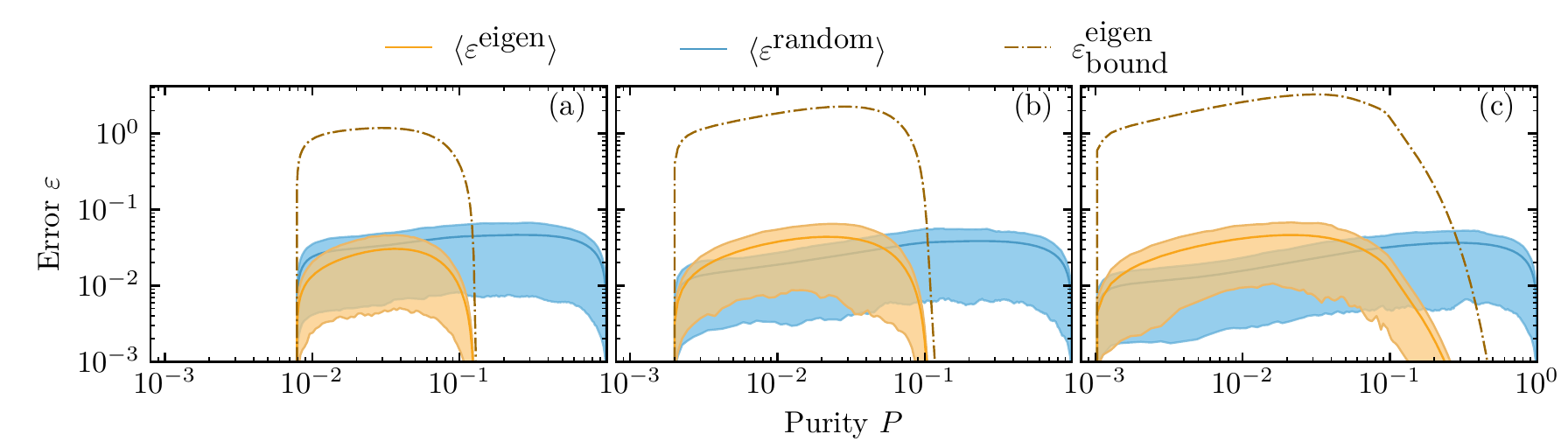}
  \caption{Average error for expectation values obtained for 200 randomized observables for a spin chain with Hilbert space dimension (a) $2^{7}=128$, (b) $2^{9}=512$, and (c) $2^{10}=1024$ using $K=10$ states in the samples. In addition to eigenstate-based sampling (orange) and the random-phase approach (blue), the analytical bound for the eigenstate-based approach according to Theorem~\ref{prop:HS_Norm} is also shown (orange, dashed).}
	\label{Fig:DimensionComparisonNew}
\end{figure*}
When increasing the number of states in the sample, $K$, the performance of both
methods improves. Most notably, the random-phase approach shows the characteristic
statistical convergence $\sim \frac{1}{\sqrt{K}}$ which is mostly independent of
ensemble purity. This behaviour is reflected by the shape of the average error,
which remains almost invariant with increasing $K$ and is merely subject to
a homogeneous downward shift. Conversely, we find for eigenstate-based sampling 
that the average error decreases far more strongly when taking larger sample 
sizes. In particular, the error peak and the performance crossing point with the
random-phase approach move towards higher purities. This behavior is intimately 
connected to the shift in the level occupations of the Boltzmann distribution. 
Since the populations decrease exponentially with increasing energy, the
eigenbasis approach can systematically reduce the error by sampling more and more
levels. In short, the random-phase approach converges homogeneously with
$\frac{1}{\sqrt{K}}$ whereas the convergence of eigenstate-based sampling is 
strongly dependent on the population distribution.

Next, we examine the performance as a function of Hilbert space 
dimension.
In Fig.~\ref{Fig:DimensionComparisonNew} we show the average error for eigenstate-based
sampling (orange) and the random-phase approach (blue) together with the
analytical bound according to Theorem~\ref{prop:HS_Norm} (orange, dashed) for
a system of 7, 9, and 10 spins (corresponding to a Hilbert space dimension of 
128, 512, and 1024, respectively).
For high purity (low temperatures), the average approximation error when using the random-phase approach appears to be roughly independent of Hilbert space dimension. For low purities (high temperatures), however, the average error actually decreases. This observation can be explained by the fact that, for large Hilbert spaces, convergence of the statistical random-phase approach is aided by the law of large numbers. For example, any two random-phase wave functions will become increasingly orthogonal when the size of Hilbert space increases, which automatically amends one of the potential issues of the random-phase approach, namely, the non-orthogonality of the sampled states. This effect is particularly pronounced at low purities which agrees with our observation that the performance of the random phase approach benefits from an increased Hilbert space dimension in this regime. As such, the gap between the two approaches closes\footnote{In fact, if $T\rightarrow\infty$ then deterministically choosing the phases of thermal wave functions such that they are elements of a mutually unbiased basis~\cite{Durt2010} with respect to the energy eigenbasis is optimal. In this case the thermal wave functions form an orthogonal set while populations still remain perfectly described by any individual wave function. In particular, if $K=N$, an exact resolution of identity is reobtained.} which reinforces the notion that the random-phase approach is particularly suitable for large and thermally hot systems.
\begin{figure*}[t]
	\centering
	\includegraphics[width=2.0\columnwidth]{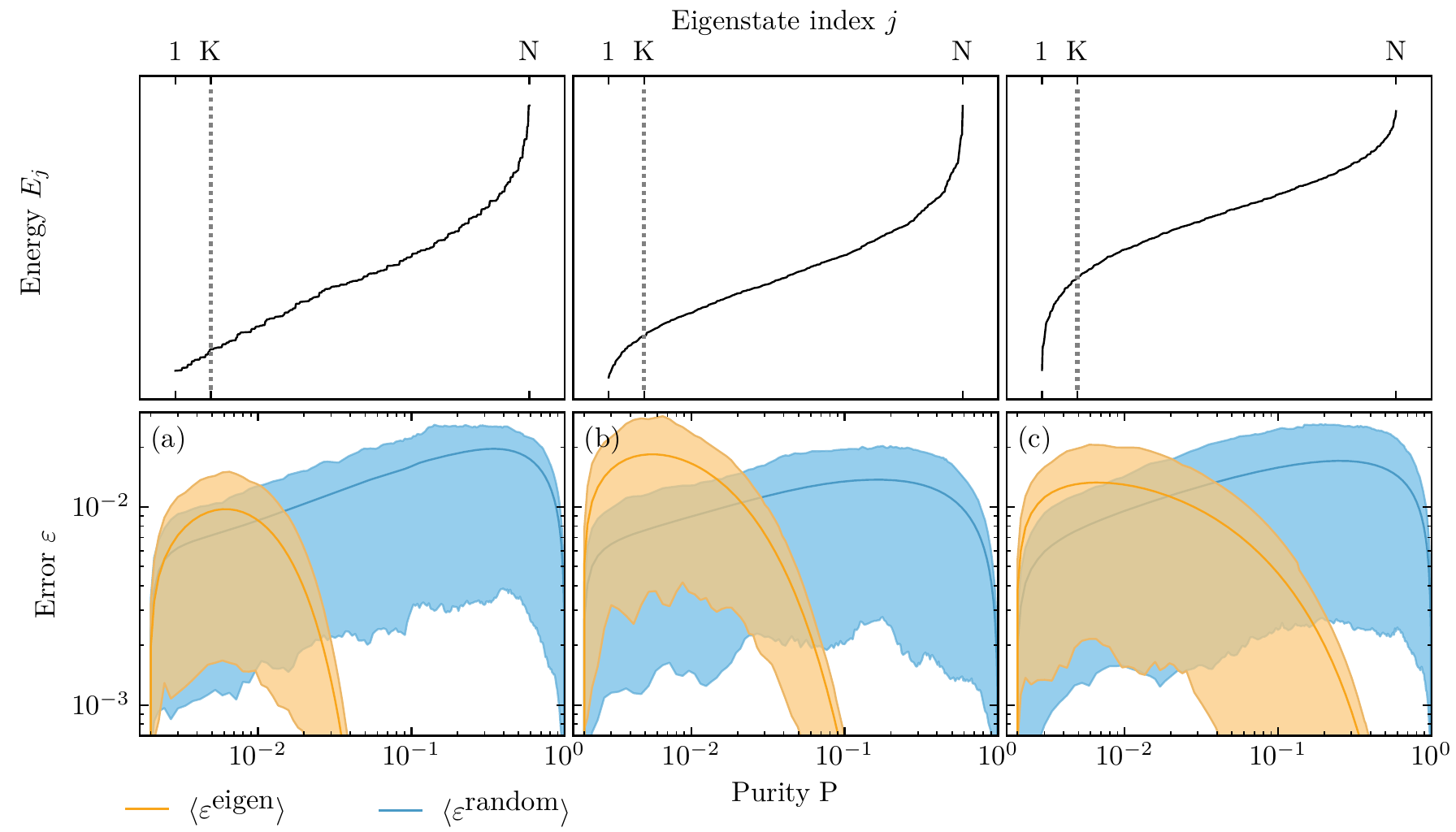}
  \caption{Average error for expectation values of 150 randomized observables in the eigenstate-based approach (orange) and using random-phase sampling (blue) for a system of 9 spins using $K=51\simeq0.10\cdot N$ sampling states. In each panel (a) - (c) the spin chain Hamiltonian was set up with a different set of parameters, thereby producing the spectra shown in the top panel. The spectral energy scale is in arbitrary units and chosen identically in all panels at the top. The vertical dotted line separates the part of the spectrum that was sampled by the eigenstate-approach to the left from the remainder to the right.}
	\label{Fig:SpectralComparison}
\end{figure*}
\begin{figure*}[t]
	\centering
	\includegraphics[width=2.0\columnwidth]{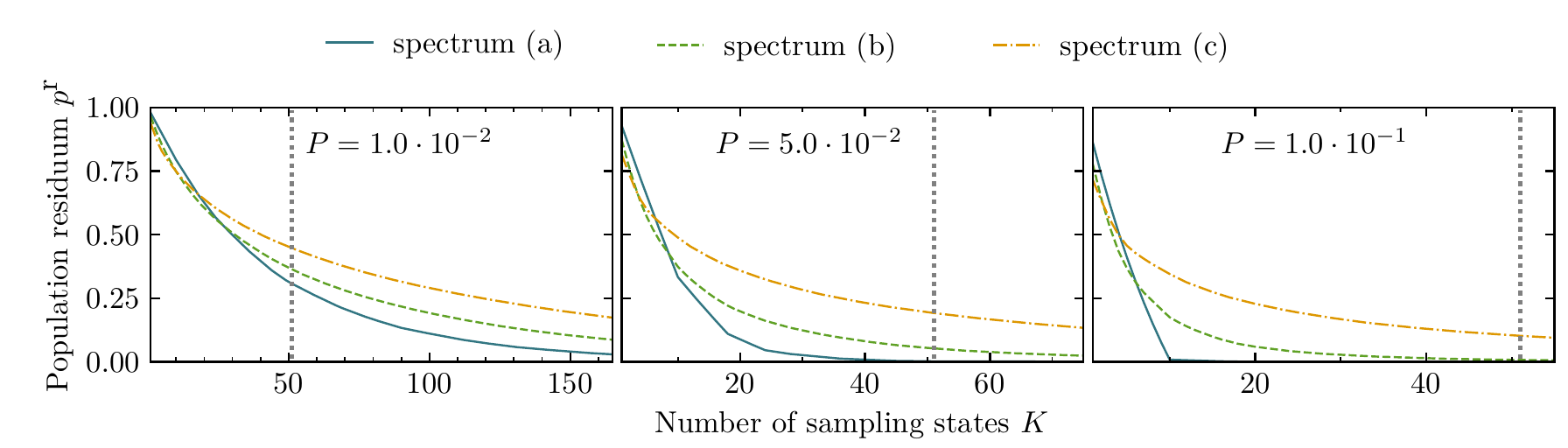}
  \caption{Population residuum $p^r(K,P)$, cf.~Eq.~\eqref{eq:pop_residuum} for $P=1\cdot10^{-2}$ (left), $P=5\cdot10^{-2}$ (center) and  $P=1\cdot10^{-1}$ (right) using identical parameters as in Fig.~\ref{Fig:SpectralComparison}. The spectra correspond to the three panels of Fig~\ref{Fig:SpectralComparison} as follows: blue curve - panel (a), green curve - panel (b), and orange curve - panel (c). The vertical dotted line fulfils the same role as in Fig.~\ref{Fig:SpectralComparison}}
	\label{Fig:SpectralComparison2}
\end{figure*}
In rather stark contrast to the random-phase approach, the performance of eigenstate-based sampling deteriorates for all purities when the Hilbert space dimension increases (cf.~Fig.~\ref{Fig:DimensionComparisonNew} (a) - (c)). 
Another particularly striking feature is given by the fact that the shape of the average error for the eigenstate-based seems to follow the shape of the worst-case error by Theorem~\ref{prop:HS_Norm} quite closely. Most notably, both curves show a very similar plateau region on the logarithmic scale at intermediate purities, which directly corresponds to the region in which eigenstate-based sampling performs at its worst compared to the random-phase approach. As a consequence, one may determine this plateau for the analytical worst-case error bound and use it as a rule of thumb to estimate the purity regions in which the random-phase approach outperforms eigenstate-based sampling. 

Our results indicate 
that the analytical error bound is
able to make profound statements on the performance of the two sampling
approaches even with respect to the average error. 
The error bound depends on the residual populations, i.e.~the populations in states which are 
not being sampled. As a consequence not only the purity can have a
large effect on the performance but also the spectrum of the Hamiltonian, or, in
other words, the density of states. In order to gain
more insight on the effect of different spectral densities, we adjusted the
parameters for the spin chain Hamiltonian in Eq.~\eqref{Eq:HamiltonianSpinChain}
to generate a set of distinct spectral shapes, cf.~the top panel of
Fig.~\ref{Fig:SpectralComparison}. An aspect of particular importance in this
context is the density of states near the ground state, since these states have
the highest population contribution to the initial ensemble and thus play a 
major role for the eigenstate-based sampling. 
The lower panels of Fig.~\ref{Fig:SpectralComparison} show the average error 
for three different spectra. The vertical dotted line shows the spectral position
of the truncation point of the sampling for the eigenstate-based approach
corresponding to $K=0.10\cdot N$ elements in the sample. From left to right, the
density of states near the ground state decreases. For the random-phase wave 
functions it is clearly visible that the form of the spectrum has barely any 
noticeable impact on the performance. Conversely, the performance of 
eigenstate-based sampling becomes noticeably worse at high purities when the 
density of states near the ground state decreases. On first glance this seems 
surprising since a low density of states implies that the population in the 
first few excited states will quickly drop off which should be beneficial for 
the eigenstate-based approach. 

In order to understand this feature we introduce the population residuum $p^r$,
\begin{equation}
p^r=1-\sum_{i=1}^{K}p_i =\sum_{i=K+1}^{N}p_i.
\label{eq:pop_residuum}
\end{equation}
This quantity reflects the population of the ensemble in states which are not taken into account in the eigenstate-based approach using a sample size of $K$ for a given purity $P$. In Fig.~\ref{Fig:SpectralComparison2} we compare the population residua as a function of $K$ for the three spectra shown in Fig.~\ref{Fig:SpectralComparison} at purities of $P=1\cdot10^{-2}$ (left), $P=5\cdot10^{-2}$ (center), and  $P=1\cdot10^{-1}$ (right). The vertical dotted line indicates the sample sizes employed to obtain the data shown in Fig.~\ref{Fig:SpectralComparison}. All three panels show that for a small number of sampled states ($K \lesssim 10$) the spectrum shown in panel (c), which yields the smallest density of states near the ground state, has the smallest population residuum compared to the other two spectra. Beyond that threshold, however, the order of the population residua inverts for the three spectra. For the sample size of $K=50$ employed in Fig.~\ref{Fig:SpectralComparison} the low density of states near the ground state actually leads to a slower convergence of the population residuum towards zero. We conclude that, roughly speaking, a low density of states near the ground state is beneficial at small sample sizes whereas for larger samples a high density of states is beneficial.

Our results for the random-phase approach qualitatively agree with the work by Kallush and Fleischer who investigated the performance of random-phase wave functions for describing orientation and alignment of a thermal ensemble of SO$_2$ molecules subject to a terahertz pulse~\cite{Kallush2015}. For high temperatures, they reported the required number of random-phase thermal wave functions to be almost independent of temperature. This is in good agreement with Fig.~\ref{Fig:ErrorImprovementComparison}, where the dependence of the error on the purity for the random-phase approach is flat for low to moderate purities. In the low-temperature (high-purity) regime they found the efficiency of the random-phase approach to deteriorate, requiring an ever-increasing amount of realizations to obtain accurate results. This matches the behavior of the average error we observe in Fig.~\ref{Fig:ErrorImprovementComparison} which does increase towards higher purities, thereby requiring a higher sample size to keep the error in check. 

Kallush and Fleischer performed simulations
investigating two specific observables, namely orientation and alignment. This leads us to the question how different the behaviour of specific observables can be from our analysis on an ``average'' observable. We will investigate this question in more detail in the following section and furthermore discuss which properties of observables can make them particularly suitable for individual sampling approaches.

\subsection{Specific Observables}
\label{subsec:ObservablesWithPriorInformation}

The different behavior we observe between average and worst-case error indicates that specific observables can behave quite differently from randomly drawn observables. In order to exemplify this, we consider the total polarization $\hat{A}_z=\sum_{j=1}^{n}\hat{\sigma}_j^z$ of the spin chain described by the Hamiltonian in Eq.~\eqref{Eq:HamiltonianSpinChain}.

\begin{figure}[t]
\centering
\includegraphics[width=\columnwidth]{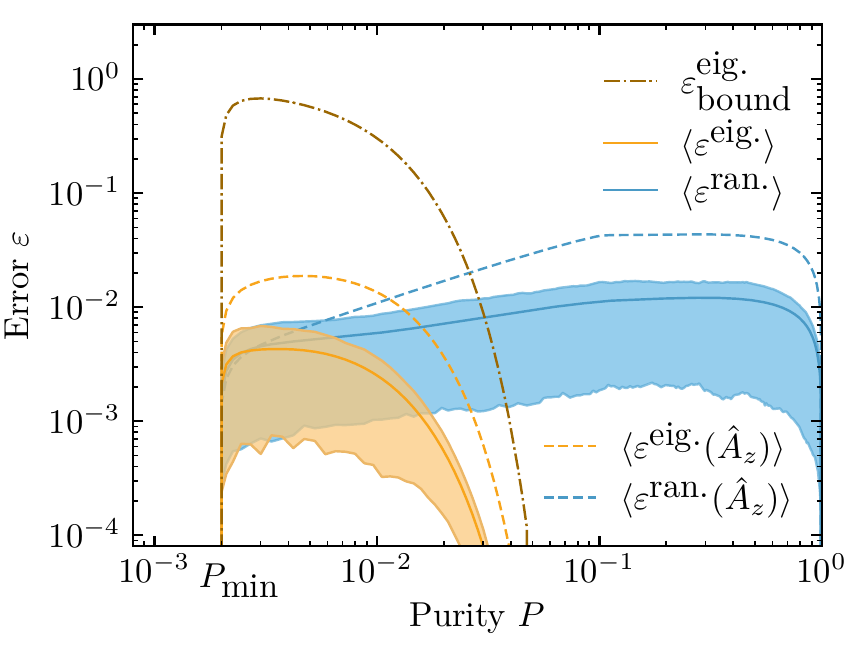}
  \caption{Average error for a set of 150 random observables (solid) and $K=100$ sampling states for a 9-spin system using eigenstate-based sampling (orange) and random-phase sampling (blue). 80\% of all errors are within the shaded area and the error bound is given for the eigenstate-based sampling (dashed-dotted). The simulations performed for obtaining the error for the $\hat{A}_z$ operator subject to 150 different fields (dashed) were performed with otherwise identical parameters to those used for the randomly drawn observables as investigated in Sec.~\ref{subsec:ArbitraryObservables}.}
\label{Fig:sum_sig_z_avg_comparison}
\end{figure}

Figure \ref{Fig:sum_sig_z_avg_comparison} shows the average error for a system of 9 spins ($N=512$) with respect to the random-phase approach (solid lines, blue), eigenstate-based sampling (solid lines, orange), and the corresponding bound according to Theorem~\ref{prop:HS_Norm} (dashed-dotted lines, orange). The solid lines, the dashed-dotted lines, and the shaded areas are analogous to those shown in Fig.~\ref{Fig:DimensionComparisonNew}. In particular, 
the shaded areas indicate the error regions into which 80\% of the performed simulations fall. The error for $\hat{A}_z$, represented by dashed lines, is averaged with respect to a set of 
randomized laser fields. All other parameters have been kept fixed with respect to the results from the previous sections, i.e.~$\hat{A}_z$ follows the same rules as all observables obtained in our randomised generation scheme. 

Quite remarkably, most of the average errors for $\hat{A}_z$ lie clearly outside the 80\% error region
obtained for randomised observables. Specifically, both sampling approaches show a worse performance at almost all purities, with the sole exception of the random-phase approach at low purities. Regarding the eigenstate-based sampling we attribute this to our observation that the diagonal matrix elements $\bra{\psi_n(t)}\hat{A}\ket{\psi_n(t)}$ in Eq.~\eqref{Eq:expectation_value_in_energy_basis} are ordered in such a way that - for the specific observable of total polarization - the first eigenstates all underestimate the true expectation value while the latter eigenstates consistently overestimate it. Therefore a large number of eigenstates needs to be sampled to compensate for the initial systematic underestimation. Despite this fact,
our analytical error bound according to Theorem~\ref{prop:HS_Norm} continues to reproduce the shape of the error curve as a function of purity rather well. This further strengthens our claim, that the error bound can be used to estimate the qualitative dependence of the average error on purity, respectively temperature. 

In order to improve our understanding of the performance of the random-phase approach, we once again move to the Heisenberg picture. Plugging Eq.~(\ref{Eq:thermal_linear_combination}) into Eq.~(\ref{Eq:expectation_value_thermal_wavefunction}) with $\ket{\phi_j}=\ket{E_j}$, i.e. constructing the random-phase wave function in the energy eigenbasis, we obtain
\begin{equation}
  \braket{\hat{A}(t)}_\beta=\lim\limits_{K\rightarrow\infty}\frac{1}{K}\sum_{k=1}^{K}\sum_{j,j'=1}^{N}\sqrt{p_jp_{j'}}e^{i(\theta_j^k-\theta_{j'}^k)}a_{j,j'}(t)\,,
\end{equation}
with $a_{j,j'}(t)=\Bra{E_{j'}}\hat{U}^\dagger(t)\hat{A}\hat{U}(t)\Ket{E_j}$. Separating the diagonal contribution yields
\begin{align}
  \braket{\hat{A}(t)}_\beta=&\lim\limits_{K\rightarrow\infty}\frac{1}{K}\sum_{k=1}^{K}\sum_{\substack{j,j'=1\nonumber\\j\neq j'}}^{N}\sqrt{p_jp_{j'}}e^{i(\theta_j^k-\theta_{j'}^k)}a_{j,j'}(t)\\
	&+\sum_{j=1}^{N}p_ja_{j,j}(t)\,. \label{Eq:diagonal_observable_separation}
\end{align}
Note that the second term in Eq.~\eqref{Eq:diagonal_observable_separation} does not depend on the random phases and thereby does not carry any sampling error. In particular, this implies that if the operator at final time in the Heisenberg picture is close to diagonal in the energy eigenbasis, then thermal wave functions are particularly suitable to sample the expectation value. This is because they only need to converge the comparatively small off-diagonal contributions, $a_{j,j'}(t)$, from the first summand in Eq.~\eqref{Eq:diagonal_observable_separation}. Generally speaking, when the observable at final time and the ensemble at initial time are close to diagonal in the same basis, i.e. $\left[\hat{A}_z(t=t_\text{final}),\hat{H}(t=0)\right]$ is small, then the random-phase approach according to Eq.~\eqref{Eq:thermal_linear_combination} is particularly suitable.

Applying this argument to our simulations of the total polarization, we note that at initial time $\hat{A}_z$ is diagonal in the energy eigenbasis, i.e., $\left[\hat{A}_z(t=0),\hat{H}(t=0)\right]=0$. Thus, ensemble and observable share an eigenbasis initially. This property will be altered by the dynamics induced by the field leading to off-diagonal contributions according to Eq.~\eqref{Eq:diagonal_observable_separation}. With respect to Fig.~\ref{Fig:sum_sig_z_avg_comparison}, this allows us to conclude that the below-average performance of the random-phase wavefunction sampling of the observable $\hat{A}_z$ is due to large off-diagonal contributions $a_{j,j'}(t)$ at final time.

\begin{figure}[t]
	\centering
	\includegraphics[width=\columnwidth]{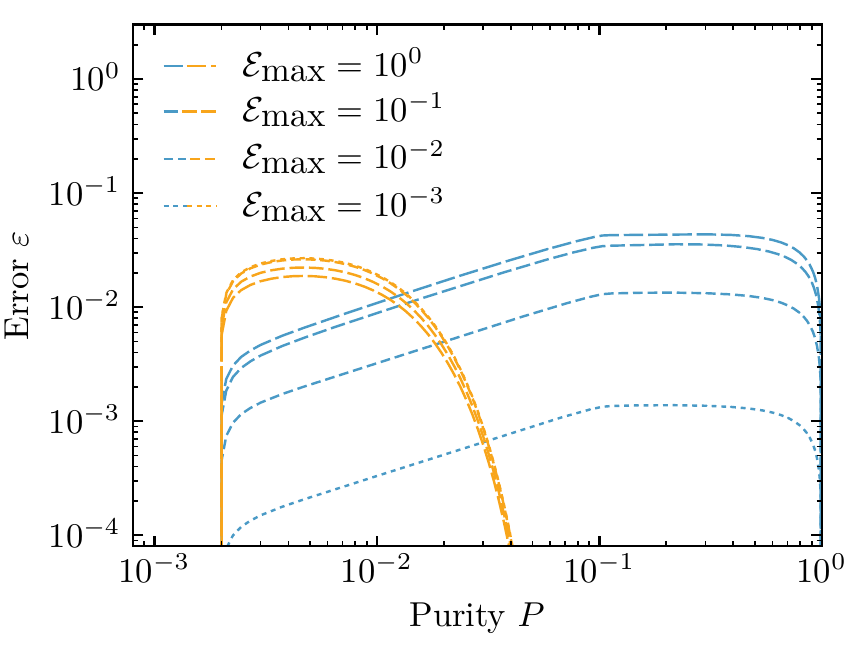}
	\caption{Average error for $\hat{A}_z$ subjected to fields with amplitudes ranging from $\mathcal{E}_\text{max}=1.0$ to $\mathcal{E}_\text{max}=0.001$ using a sample size of $K=100$ in a 9-spin system for eigenstate-based sampling (orange) and the random-phase approach (blue). A longer dash length represent higher field amplitude.}
	\label{Fig:sum_sig_z_field_scaling}
\end{figure}

In order to verify that these contributions are caused by the introduction of off-diagonal contributions during to time evolution, we show in Fig.~\ref{Fig:sum_sig_z_field_scaling} the average error for both sampling methods for different field amplitudes, ranging from $\mathcal{E}_\text{max}=0.001$ to $\mathcal{E}_\text{max}=1.0$ with all other simulation parameters identical to Fig.~\ref{Fig:sum_sig_z_avg_comparison}. 
We argue that the sensitivity of the random-phase approach's performance observed in Fig.~\ref{Fig:sum_sig_z_field_scaling} is due to the fact that strong fields are more likely to destroys a significant portion of the diagonal structure of the total polarization in the energy eigenbasis. Conversely, for eigenstate-based sampling, we observe only a minor dependency on $\mathcal{E}_\text{max}$. 

In summary, the structure of the observable in the Heisenberg picture can have a major impact on the convergence behavior, specifically regarding the magnitude of its off-diagonal matrix elements in the ensemble eigenbasis. Unfortunately, the magnitude of these elements is typically hard to estimate, since it requires full diagonalization to obtain all eigenvectors. In addition, the dynamics will shift the balance between diagonal and off-diagonal elements. However, such effects need not be solely detrimental. Rather, they can also be systematically exploited as we will demonstrate in the following.

\section{Optimal sampling for low rank observables}
\label{sec:opt_sampling_low_rank_observable}

For density matrices with high purity, such as thermal states at low temperatures, it is sufficient to propagate only few energy eigenstates, as evidenced by Eq.~\eqref{Eq:expectation_value_in_energy_basis}. For ensembles with low purity, to keep the error in check the required fraction of states for eigenstate-based sampling can become quite large, cf.~Theorem~\ref{prop:HS_Norm}. Furthermore, the structure of the observable in conjunction with the structure of the initial density matrix can play a major role, as evidenced by our findings in Sec.~\ref{subsec:ObservablesWithPriorInformation}. We show in the following that, by swapping the treatment of observable and initial density matrix for eigenstate-based sampling, we are able to apply Theorem~\ref{prop:HS_Norm} to exploit structural effects on the level of the observable. Specifically, we devise an efficient sampling scheme for low-rank observables, which are rather naturally obtained when considering projectors. For example, simulations of thermal systems that require the measurement of populations in a set of bound states on specific electronic surfaces~\cite{AmaranJCP2013,LevinPRL2015} naturally give rise to such operators. The projector rank is equal to the number of bound states, which is typically much smaller than the total Hilbert space dimension.

We make use of the fact that the roles of initial density matrix and observable are interchangeable when calculating expectation values due to cyclic invariance of the trace,
\begin{equation}
\text{tr}[\hat{A}\hat{U}(t)\hat{\rho}_\beta\hat{U}^\dagger(t)]=\text{tr}[\hat{U}^\dagger(t)\hat{A}\hat{U}(t)\hat{\rho}_\beta]\,.
\end{equation}
Using this relation, expectation values can be calculated via
\begin{eqnarray}
\braket{\hat{A}}_\beta(t)&=&\text{tr}[\hat{U}^\dagger(t)\hat{A}\hat{U}(t)\hat{\rho}_\beta]\nonumber\\
&=&\sum_{n,m=1}^{N}\bra{A_{m}}\hat{U}^\dagger(t) a_n\ket{A_n}\bra{A_n}\hat{U}(t)\hat{\rho}_\beta\ket{A_{m}}\nonumber\\
&=&\sum_{n,m=1}^{N}a_n\bra{A_n}\hat{U}(t)\hat{\rho}_\beta\ket{A_{m}}\bra{A_{m}}\hat{U}^\dagger(t)\ket{A_n}\nonumber\\
&=&\sum_{n=1}^{N}a_n\bra{A_n}\hat{U}(t)\hat{\rho}_\beta\underbrace{\hat{U}^\dagger(t)\ket{A_n}}_{\ket{A_n(t)}}\nonumber\\
&=&\sum_{n=1}^{K}a_n\bra{A_n(t)}\hat{\rho}_\beta\ket{A_n(t)}\label{Eq:expectation_value_in_operator_basis}\,,
\end{eqnarray}
where $\hat{A}=\sum_{n=1}^{N}a_n\ket{A_n}\bra{A_n}$ with eigenvalues $\{a_n\}_{n=1,\dots,N}$ and eigenvectors $\{\ket{A_n}\}_{n=1,\dots,N}$. Note that each $\ket{A_n(t)}=\hat{U}^\dagger(t)\ket{A_n}$ can be interpreted as a backwards-propagated eigenstate of the observable. The low rank of $\hat{A}$ is important since any particular propagated eigenstate $\ket{A_n(t)}$ will not contribute to the sum in Eq.~\eqref{Eq:expectation_value_in_operator_basis} if the corresponding eigenvalue $a_n$ is zero. Since low rank operators possess a large number of vanishing eigenvalues, only a small amount of states need to be propagated. More generally, propagating the first $K$ eigenstates corresponding to the $K$ eigenvalues with the largest modulus leads to the following approximation of the observable $\hat{A}$ (represented in its eigenbasis),
\begin{equation}
(\hat{A}_\text{Approx})_{ij}=\begin{cases} a_i &\text{if } i=j<K, \\
0 & \text{otherwise}, \end{cases}
\end{equation}
which, by Theorem~\ref{prop:HS_Norm}, yields the smallest attainable worst-case error bound for arbitrary initial ensembles of
\begin{equation}
\varepsilon\leq\sqrt{\sum_{i=K+1}^{N}\lvert a_i\rvert^2}\label{Eq:sum_lowest_eigenvalues_obs}\,,
\end{equation}
with $\varepsilon$ being defined as in Eq.~\eqref{Eq:TrueError}. In particular, if $\hat{A}$ is of rank $K$ and the corresponding $K$ eigenstates are employed, then $\hat{A}_\text{Approx}=\hat{A}$ and therefore $\varepsilon$ will vanish, too. Note that the error bound does not depend on the system dynamics or on any features of the initial ensemble. In particular, the performance is not negatively affected by high temperatures / low purity.

To improve the sampling performance, it is even possible to use the enhancements introduced in Sec.~\ref{sec:sampling_enhancements} for the observable sampling presented in this section, i.e.~convergence can be improved by removing the identity component, respectively the trace, of the density matrix, respectively the observable. The only drawback of observable-based sampling is that the set of states that need to be propagated is tailored to the specific observable. Even though expectation values for arbitrary initial states can be approximated with the resulting set of propagated states, repeating the simulation for a different observable would require, in the worst case, the propagation of a completely different set of states. In short, sampling the observable is appropriate if one is interested in a single physical quantity for different initial states whereas sampling different observables for the same initial state is more suitably performed with the approach from Sec.~\ref{sec:opt_sampling_arb_observable}.

\section{Conclusions}
\label{sec:conclusions}

We have considered the approximation of time-dependent observables in a quantum system described by a statistical ensemble undergoing coherent time evolution. 
We have shown that, with respect to minimizing the worst-case error, there is an optimal approach to compute arbitrary time-dependent observables via pure-state sampling. It consists of using the lowest-lying energy eigenstates. The corresponding error is bounded by the summed eigenvalues of those eigenstates which are not included in the sampling. Eigenstate-based sampling is the uniquely optimal choice in this case. In particular, the worst-case sampling error is smaller than in any randomized sampling approach. Nevertheless the performance regarding the \textit{average} error in a particular system for a set of observables is only superior to random-phase sampling if the ensemble purity is relatively high or, in the language of thermal systems, the ensemble is cold. This can be attributed to the fact that eigenstate-based sampling is constructed hierarchically, starting from the energetically lowest-lying states. Thus, it might easily miss important contribution from high-lying states which start to play an important role once mixedness increases. For low purities, a randomized approach provides on average a much more suitable coverage of Hilbert space.

While our analytical error bound, cf.~Theorem~\ref{prop:HS_Norm}, is not suitable to quantitatively estimate the average error of the eigenstate-based approach, it still allows to make predictions on the general dependence of its performance on, e.g., ensemble purity. We found that the spectrum of the initial ensemble, which for thermal systems is directly correlated with the density of states, has a noticeable impact on the sampling performance - a low density of states near the ground state coincides with favourable scaling at small sample sizes, whereas a high density of states is beneficial for larger samples. Furthermore, we found that it is still possible to make concrete qualitative statements on the eigenstate-based approach's peformance from the analytical bound in a wide variety of situations, even when the average error instead of the worst-case error is considered.

Furthermore we have shown that convergence of the two methods can be accelerated
by avoiding to include 
quantities in the sampling which are known \textit{a priori}. This can be achieved by removing the identity component of both the observable and the density matrix. 
This implies that sampling should always be performed with traceless observables. While for eigenstate-based sampling, both enhancements proved to be generally
beneficial, the random-phase approach benefits from these corrections primarily at low ensemble purities. In the limit of very low purities, the removal of the identity components can reduce the average error by several orders of magnitude.

If one is only interested in specific physical quantities, then prior information about the structure of the corresponding observables can be exploited to refine the sampling scheme. Most notably, if the rank of the observable of interest is low, it is possible to propagate the eigenstates corresponding to the largest eigenvalues of the observable of interest backwards in time. Then, one can compute the overlap of this backwards-propagated observable with the initial state at $t=0$ to obtain the corresponding expectation value. This works particularly well for low-dimensional projectors. For the special case of one-dimensional projectors, it implies that propagation of a single wave function is sufficient to compute the expectation values for arbitrary ensembles.

Although our notation and our examples were inspired by thermal states, we have made no assumptions on the particular physical nature of the initial state. Our main theorem and the methods of random-phase and eigenstate-based sampling can be employed for arbitrary density matrices. The relevant eigenstates and eigenenergies in the general case are those of the initial density matrix instead of the Hamiltonian. We therefore expect our findings to be applicable not only to thermal states but also to different types of mixed states as encountered, for example, in mixed-state quantum computing~\cite{Ambainis2006,Shor2008}. 

Finally, our results regarding the use of prior information on the observable or initial state immediately raise the question whether it is also possible to exploit prior information on the system dynamics. For example, one could make use of the fact that all mixed states possess no coherences in their eigenbasis. If the evolution does not introduce significant coherences and the observable is diagonal in this basis, then a single Hilbert space state is sufficient as long as it correctly reproduces the populations at $t=0$. Saving numerical effort by adaptively focusing on only the most relevant part of Hilbert space is also at the core of many modern approaches in quantum chemistry, see, e.g., Refs.~\cite{Schriber2016,Tubman2016}, and condensed matter physics, see, e.g., Refs.~\cite{Gualdi2013,Go2017}. We expect that gaining more insight into the question of how to best perform such truncations will open up further avenues to fight the curse of dimensionality in quantum dynamics.

\begin{acknowledgments}
Financial support by the Deutsche Forschungsgemeinschaft via the priority program SPP1840 QUTIF is gratefully acknowledged. We would like to thank Gustavo M. Pastor for many fruitful discussions as well as 
all referees for their insightful comments which have helped us to substantially
improve this work.
\end{acknowledgments}

\end{document}